\documentclass[times,twocolumn,final,authoryear]{elsarticle}

\usepackage{jasr}
\usepackage{framed,multirow}

\usepackage{amssymb}
\usepackage{latexsym}

\usepackage{url}
\usepackage{xcolor}
\definecolor{newcolor}{rgb}{.8,.349,.1}

\usepackage[citebordercolor=white]{hyperref}
\usepackage{comment}

\journal{Advances in Space Research}

\begin{document}

\verso{Paul Chote \textit{et. al.}}

\begin{frontmatter}
    \title{High-precision light curves of Geostationary objects:\\The \textit{PHANTOM ECHOES 2} RPO campaign}

    \author[Warwick-astro,Warwick-csda]{Paul \snm{Chote}\corref{cor}}
    \cortext[cor]{Corresponding author}
    \ead{p.chote@warwick.ac.uk}
    \author[Warwick-astro,Warwick-csda]{Robert \snm{Airey}}
    \author[Warwick-astro,Warwick-csda]{James \snm{McCormac}}
    \author[Warwick-astro,Warwick-csda]{Don \snm{Pollacco}}
    \author[Warwick-astro,Warwick-csda]{Richard \snm{West}}
    \author[Warwick-astro]{Krzysztof \snm{Ulaczyk}}
    \author[Sheffield]{Martin J. \snm{Dyer}}
    \author[Dstl-Portsdown]{Alexander \snm{Agathanggelou}}
    \author[Dstl-Portsdown]{William \snm{Feline}}
    \author[Dstl-Portsdown]{Simon \snm{George}}
    \author[Dstl-PortonDown]{Calum \snm{Meredith}}
    \author[Dstl-PortonDown]{Grant \snm{Privett}}

    \address[Warwick-astro]{Department of Physics, University of Warwick, Coventry, CV4 7AL (UK)}
    \address[Warwick-csda]{Centre for Space Domain Awareness, University of Warwick, Coventry, CV4 7AL (UK)}
    \address[Sheffield]{Astrophysics Research Cluster, School of Mathematical and Physical Sciences, University of Sheffield, Sheffield, S3 7RH (UK)}
    \address[Dstl-Portsdown]{Defence Science \& Technology Laboratory, Portsdown West, Fareham, PO17 6AD (UK)}
    \address[Dstl-PortonDown]{Defence Science \& Technology Laboratory, Porton Down, Salisbury, SP4 0JQ (UK)}

    \received{14 Feb 2025}
    \finalform{16 April 2025}
    \accepted{8 May 2025}

\begin{abstract}

We present results from an extensive optical observation campaign that monitored the Geostationary satellites Intelsat 10-02, Mission Extension Vehicle 2, Thor 5, Thor 6, Thor 7, and Meteosat 11 over a 14 week period that covered the proximity operations and docking of Mission Extension Vehicle 2 with Intelsat 10-02. High-cadence single-color photometric observations are supplemented with targeted multi-color observations, high resolution imaging, and passive radio frequency positioning obtained using complementary facilities.

The photometric signatures of the six targets are presented in the form of two-dimensional color maps. A selection of interesting features are investigated in further detail, including a rapid glinting behavior in Thor 6; a brightening event from Meteosat 11; using glints to constrain the unresolved positions of Intelsat 10-02 and MEV-2; changes in the photometric signature of Intelsat 10-02 before and after docking; and signatures of attitude changes and maneuvering in the light curves of MEV-2.

A detailed description of the photometric data reduction pipeline is also presented, with a focus on details that must be considered when aiming for sub-percent photometric precision.

\end{abstract}

\end{frontmatter}

\section{Introduction}
\label{sec:introduction}

The commercial development of space is undergoing a phenomenal expansion, driven by dramatic decreases in launch costs. Most of this development is focused on low Earth orbits (LEO), but the Geostationary (GEO) orbits at an altitude of $\approx36000\,$km remain of critical importance due to their unique orbital geometry: satellites (resident space objects, RSOs) at this altitude have a 24\,h orbital period, which allows them to point continuously at a fixed footprint on the ground with minimal fuel expenditure. This makes them well suited for communication and broadcast services, and to global weather monitoring.

The GEO environment requires particular care, as there are no natural sinks (c.f. atmospheric drag at LEO) to remove debris \citep{esa2023}. As the space sector continues to grow, so does the need to demonstrate robust and cost-effective techniques for maintaining an active situational awareness. While there are promising developments towards satellite-mounted hardware that enhances trackability \citep[e.g.][]{Krag2024, Bakker2024}, these solutions focus on small LEO satellites and have not yet been deployed in significant numbers.

Traditionally, remote observations of GEO RSOs were limited to simply maintaining custody of large objects by tracking their position and mean brightness \citep[e.g.][]{Schildknecht2001, Alby2004, Barker2005}. In more recent years, interest has moved towards using measurements of RSO brightness over time (called a `light curve') to infer details about their nature and behavior without the need for resolved imagery \citep[e.g.][]{Schildknecht2008, Papushev2009, Cognion2014, Cardona2016, Kelecy2022, Mariani2023}.

GEO RSOs are currently treated as disposable objects: they can only hold a limited supply of propellant for station-keeping, and the performance of their solar panels degrades over time. Operators are now required to raise the orbit of their RSOs at the end of their life into a so-called 'graveyard' orbit above the protected GEO region, but around 20\% of end-of-life RSOs are still failing to achieve this \citep{esa2023}.

One intriguing area that is currently being developed by commercial operators is on-orbit servicing, where a dedicated service vehicle is launched to replenish, repair, or upgrade a client vehicle -- either by attaching itself to the client to replace capabilities, or by directly manipulating the client using robotic arms and tools.

The first commercial servicing mission occurred in February 2020, with the docking of Mission Extension Vehicle 1 (MEV-1) with Intelsat 901. The maneuvers for this docking were undertaken in a graveyard orbit to reduce the debris risk to the GEO environment if a catastrophic failure occurred -- docking requires the two RSOs to cooperatively operate in extremely close proximity without colliding -- before the combined vehicle returned under the propulsion of MEV-1 to its designated orbital slot.

Ground-based observations of these rendezvous and proximity operations (RPOs) are challenging because the close physical separations make it difficult to individually resolve the separate RSOs. The \textit{PHANTOM ECHOES} project \citep{George2020} was undertaken to examine how ground- and space-based observations can be usefully applied to monitor these scenarios using a real-world cooperative test-case.

A second Mission Extension Vehicle (aptly called MEV-2) launched in August 2020 with the ambitious goal of docking with Intelsat 10-02 while still in GEO, without interrupting its broadcast service. MEV-2 reached Intelsat 10-02 in late January 2021, and successfully docked on April 12 after several weeks of RPO.

\textit{PHANTOM ECHOES} continued over this second mission \citep{George2021}, and had a particular focus on the RPO period due to its near-ideal observability from European facilities. Extensive ground-based observations were obtained for six RSOs:

\begin{itemize}
\item \textbf{Intelsat 10-02} (IS10-02; SATCAT \#28358) is a communications satellite launched in 2004 to supply digital broadcast, telephone, and internet services \citep{KrebsIntelsat}. It is based on Airbus's Eurostar 3000 satellite bus.
\item \textbf{Mission Extension Vehicle 2} (MEV-2; SATCAT \#46113) was developed by Space Logistics LLC (a subsidiary of Northrop Grumman) with the design goal of extending the life of commercial GEO spacecraft by clamping on to the nozzle of the liquid apogee engine present on most GEO RSOs. The MEV would then take over all station-keeping and attitude control maneuvers \citep{Pyrak2021}, and deposit the client in a graveyard orbit at the end of its mission. It is based on a modified version of the GEOStar bus, and has an expected mission lifetime of 15 years -- sufficient to service multiple client RSOs.
\item \textbf{Thor 5} (SATCAT \#32487) provides satellite and broadcast services for Telenor Satellite Broadcasting AS. It was built by Orbital Sciences Corporation on their GEOStar-2 platform, and features a pair of deployable 2.3m reflectors plus a steerable 0.75m antenna. It was launched in 2008 with an expected lifetime of 15 years \citep{WebThor5}.
\item \textbf{Thor 6} (SATCAT \#36033) broadcasts satellite TV for Telenor Satellite Broadcasting AS. It is based on Thales Alenia's Spacebus 4000B2 platform, and was launched in 2009, also with a 15 year lifetime \citep{WebThor6}.
\item \textbf{Thor 7} (SATCAT \#40613) provides additional broadcast TV capacity for Europe and serves internet connectivity through a combination of fixed spot-beams and a steerable spot beam \citep{WebThor7}. It was manufactured by Space Systems Loral using their SSL-1300 bus.
\item \textbf{Meteosat 11} (SATCAT \#40732) is a second-generation Meteosat satellite designed for EUMETSAT by the European Space Agency \citep{WebMeteosat}. It has a drum-shaped body that spins at 100\,RPM, scanning across the disk of the Earth to obtain multi-wavelength imagery for weather forecasting. It was launched in 2015 with an anticipated operational lifetime of 7 years.
\end{itemize}

RSOs do not intrinsically emit at optical wavelengths; the flux that we measure comes from sunlight reflected from their surfaces. It should therefore be unsurprising that the behavior of the light curves is driven primarily by changes in the geometry of the Sun, the RSO, and the observer.

This geometry is best captured by the concept of the solar phase angle, and more specifically by the solar \emph{equatorial} phase angle \citep{Payne2007}. While the standard phase angle measures the angle bisecting the Sun-RSO and RSO-Observer vectors, the equatorial phase angle corresponds specifically to its signed longitudinal component -- or equivalently, the difference in the apparent Right Ascension of the RSO and the anti-solar point as seen by an observer.

The equatorial phase angle (hereafter just `phase angle') is a useful coordinate against which to plot light curves because it naturally describes the illumination geometry of the RSO. A phase angle of $0^\circ$ always corresponds to when the Sun is directly behind the observer relative to the RSO. This removes a degeneracy between UTC time and RSO longitude, making it simpler to compare light curves of RSOs that are at different positions on the sky. This coordinate system also makes it easy to identify the impact of the Earth's shadow, which for a few weeks (centered on the equinoxes) sweeps across the GEO belt over the course of the night, but is always near $0^\circ$ phase (it will be offset from zero due to parallax caused by the observer being offset from the geocenter).

Many GEO RSOs feature a box-wing configuration, with large flat solar panel arrays extending from the northern and southern faces of a rectangular body and pivoting around this axis to track the Sun. Satellites of this type typically show a strong `glint' (brightness peak) near $0^\circ$ phase caused by an increase in specular reflection from the solar panels as the angles of incidence and reflection approach zero. Satellite operators may choose to deliberately offset the solar arrays away from the Sun for operational reasons (e.g. to match power generation to demand, or to balance torques from solar radiation pressure); these offsets can be measured directly as half of the measured glint peak phase angle \citep{Payne2006}.

Light curves also show seasonal variation due to the changing declination of the Sun; in particular, the `glint season' around the solar equinoxes when the Sun-RSO-Observer plane aligns with the rotation axis of the Earth, maximizing the specular reflection from the solar panels.

Other features in the light curves can be associated with reflections off the satellite bus, which typically holds a fixed orientation relative to the Earth (in order to maintain a static antenna footprint). These features may drift in equatorial phase angle as a function of the solar declination \citep{Hall2013}.

The amount of reflected light is generally not uniform with wavelength, so RSOs made with different materials may show distinctive color relationships that could be used to uniquely distinguish them \citep[e.g.][]{Schmitt2020, Yu2022}. In particular, changes in color during glints may provide insight into which part of the RSO is causing the reflection \citep[e.g.][]{Hall2010, Zilkova2023, Zigo2023}.

Here we report on a significant body of optical photometry obtained using three facilities during the \textit{PHANTOM ECHOES 2} RPO campaign. The observations are described in Section \ref{sec:observations}, followed by a detailed description of the data reduction pipeline in Section \ref{sec:reduction} with an emphasis on the corrections required to attain $\sim$ percent-level uncertainties. Section \ref{sec:results} presents a broad overview of the data sets, with Section \ref{sec:selected-cases} providing a deeper analysis of selected features. Finally, Section \ref{sec:conclusions} attempts to contextualize the results and discusses the future steps for our observational program.

\section{Observations}
\label{sec:observations}

\subsection{Warwick Test Telescope}
\label{sec:observations-cmos}

The majority of observations were obtained with the University of Warwick's test telescope (WTT) installation at the Roque de los Muchachos observatory on La Palma in the Canary Islands. At the time of these observations, this facility consisted of a co-pointed pair of 18cm Takahashi Epsilon 180-ED astrographs repurposed from the \textit{Next Generation Transit Survey} (NGTS) prototype \citep{2018McCormac, Wheatley2018} on an equatorial Paramount ME mount.

Only one of the astrographs was used for these observations, which was fitted with an Andor Marana scientific CMOS detector. The GSense 400 sensor in the Marana has $2048 \times 2048$ 11 $\mu$m pixels, giving a field of view of $2.6^\circ \times 2.6^\circ$ with a plate scale of $4.5''/$pixel. In order to maximize throughput, the system operated without a filter. This wide field of view allowed several RSOs to be simultaneously monitored: IS10-02 and MEV-2 themselves, plus Thor 5, Thor 6, Thor 7, and Meteosat~11.

Observations were obtained robotically over a 14-week period, with the telescope tracking disabled so that GEO RSOs appeared as point sources on the detector and background stars appeared as short horizontal streaks. A baseline exposure time of 5 seconds was chosen as a compromise between resolving short-timescale behaviors and keeping a manageable data rate. The Marana uses a rolling electronic shutter, which provides effectively zero read-out dead time between exposures but introduces a time skew as a function of detector row that must be accounted for during image reduction.

Because the observation campaign spanned glint season, the brightness of a given target could change by as much as 6 magnitudes over the course of the night. The time ranges where one or more of the targets was predicted to saturate was calculated each night, and the exposure time reduced to 1 second during these periods. The full observation log is listed in Table \ref{tab:cmos-observations}.

\subsection{GOTO-North}
\label{sec:observations-goto}

Additional supporting observations were obtained using the prototype installation of the GOTO-North facility \citep{Steeghs2022} to obtain simultaneous 2- and 3-color photometry of the RSOs. Each GOTO mount consists of an array of eight 40\,cm telescopes that are mosaiced together to form a larger continuous field of view. The unit telescopes use FLI Microline cameras with the KAF-50100 CCD sensors. The main GOTO survey operates with a broad optical, UV/IR cut filter (Baader L band), but the unit telescopes are equipped with a filter wheel that contains photographic red (R), green (G), and blue (B) filters (wavelength ranges are provided in Table \ref{tab:goto-calibration}).

\begin{figure}[htbp]
    \centering
    \includegraphics[width=0.4\textwidth]{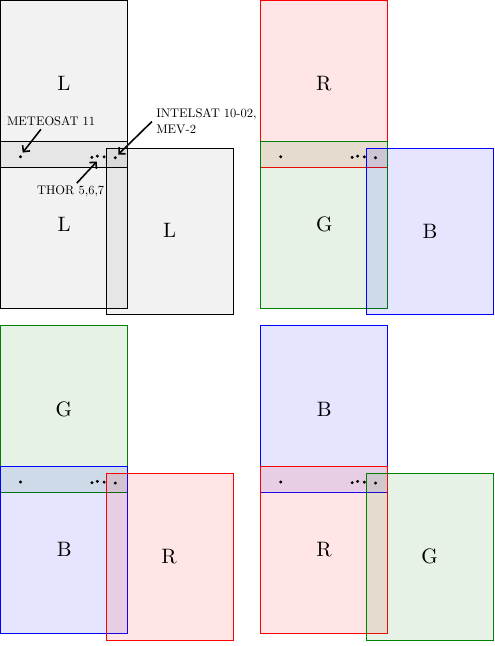}
    \caption{Schematic of the GOTO-North observation sequence showing the overlapping field of view of three unit telescopes. Groups of 5 $\times$ 5 second exposures were taken, cycling filter combination between each group to measure simultaneous pairs of colors.}
    \label{fig:gotofilters}
\end{figure}

The GOTO prototype was configured with relatively large overlap between the central unit telescopes, which made it possible to position the target RSOs in the overlap regions where IS10-02 and MEV-2 could be simultaneously observed in three unit telescopes, and the neighboring Thor 5, 6, 7 and Meteosat satellites in two unit telescopes. Figure \ref{fig:gotofilters} illustrates how, by cycling filters between the unit telescopes, simultaneous observations in each filter pair could be obtained to measure high-precision color indices.

Table \ref{tab:goto-observations} lists the observations, which consisted of several sets of $5 \times 5$ second exposures. Useful observations were obtained on 21 nights, with most observation sequences starting during the twilight gap between flat-field calibrations and GOTO's primary science program. An additional two observations covered different phase angle ranges to investigate the color behavior of glints in IS10-02 and Thor 6.

\subsection{SAFRAN WeTrack\texttrademark}
\label{sec:observations-safran}

SAFRAN Data Systems were contracted through the PHANTOM ECHOES project to supply data from their commercial WeTrack\texttrademark~service covering the span of the RPO campaign.

WeTrack uses a network of radio antennas positioned across France to passively measure the radio frequency (RF) signals emitted by active RSOs. The signals measured from pairs of antennas are correlated to produce time-difference of arrival (TDOA) and frequency-difference of arrival (FDOA) measurements at a regular cadence across multiple baselines. Multilateration using these measurements can be performed in order to measure the instantaneous position and velocity of the emitter; although, in practice, several days of measurements are fitted to produce an orbital solution that is interpolated to a 1 minute cadence for distribution (B. Guillot, private comm). 

Data were provided in the form of CCSDS OEM files \citep{CCSDS5020B3}, which contain three-dimensional positions and velocities in the ITRF2000 coordinate system for IS10-02, MEV-2, Thor 5, Thor 6, and Thor 7.

\section{Data Reduction}
\label{sec:reduction}

The optical data were processed using a custom data reduction pipeline designed for extracting calibrated photometry from non-sidereal images. The pipeline, first described in \cite{Chote2019}, has since undergone significant enhancements that warrants a complete description below.

Data are processed through several steps, starting with frame calibration, followed by source detection, track fitting, and finally light curve extraction. The pipeline has been developed in Python and uses several high-quality open source community packages: Astropy \citep{astropy:2013, astropy:2018}, SEP \citep{Barbary2016, Bertin1996}, NumPy \citep{Harris2020}, SciPy \citep{2020SciPy}, Skyfield \citep{skyfield}, and photutils \citep{photutils}.

\subsection{Frame Calibrations}
\label{sec:reduction-calibration}

Standard bias and flat field corrections are applied, and the background sky level is fitted and subtracted using a two dimensional map. Sources are detected using a filter matched to the expected trail length for background stars, and the detected centroid positions are given to Astrometry.net \citep{Lang2010} to obtain an initial world coordinate system (WCS) astrometric solution for the mid-exposure frame time.

\begin{figure}[htbp]
    \centering
    \includegraphics{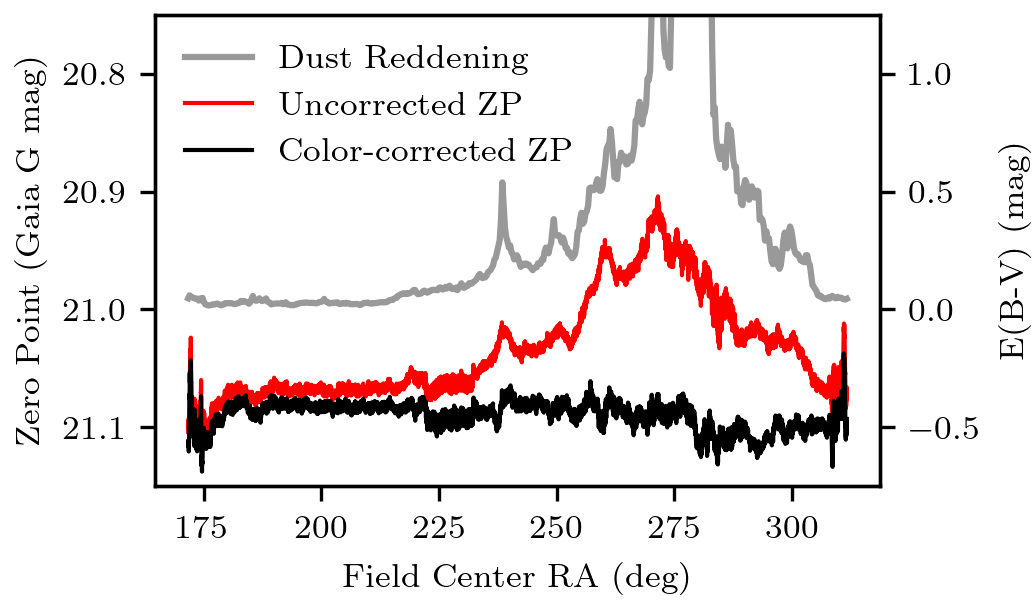}
    \caption{A plot of zero point as a function of field center RA, obtained by equating WTT instrumental magnitudes directly against the Gaia G bandpass (top) on a clear night shows a strong correlation with the \cite{Schlafly2011} galactic dust reddening (via the NASA/IPAC infrared science archive). Including a linear term in G\textsubscript{BP} - G\textsubscript{RP} when calculating the zero point eliminates this and reduces the overall systematics to below 0.05 mag (bottom). In both cases, the standard deviation of the point-to-point scatter (measured in the flat region between RA $200 - 210^\circ$ to avoid the dust systematics) is below 0.003~mag.}
    \label{fig:dustreddening}
\end{figure}

WTT observations are calibrated against the Gaia DR2 catalog \citep{GaiaCollaboration2018}, matching instrumental fluxes to the G band-pass with a G\textsubscript{BP}~-~G\textsubscript{RP} color-correction term. The frame zero point is evaluated for (G\textsubscript{BP} - G\textsubscript{RP})$_\odot$ = 0.82, the Gaia color of the Sun \citep{Casagrande2018}. This color correction is needed to account for the small differences between the instrumental and Gaia G bandpasses; without this, the effect of galactic dust reddening introduces a systematic error of a few tenths of a magnitude as shown in Figure \ref{fig:dustreddening}.

The calibration catalogs are pre-processed to add a column that identifies whether a star is a suitable calibration reference based on:

\begin{enumerate}
    \item Photometrically non-variable.
    \item Has measurements for all of G, G\textsubscript{RP}, and G\textsubscript{BP} magnitudes.
    \item Brighter than 14th magnitude in G.
    \item Has no neighbors brighter or within 2.5 magnitudes fainter in a ``blending box'' with sides $\Delta$RA = 120 arcsec and $\Delta$Dec~=~45 arcsec.
\end{enumerate}

\begin{figure}[htbp]
    \centering
    \includegraphics{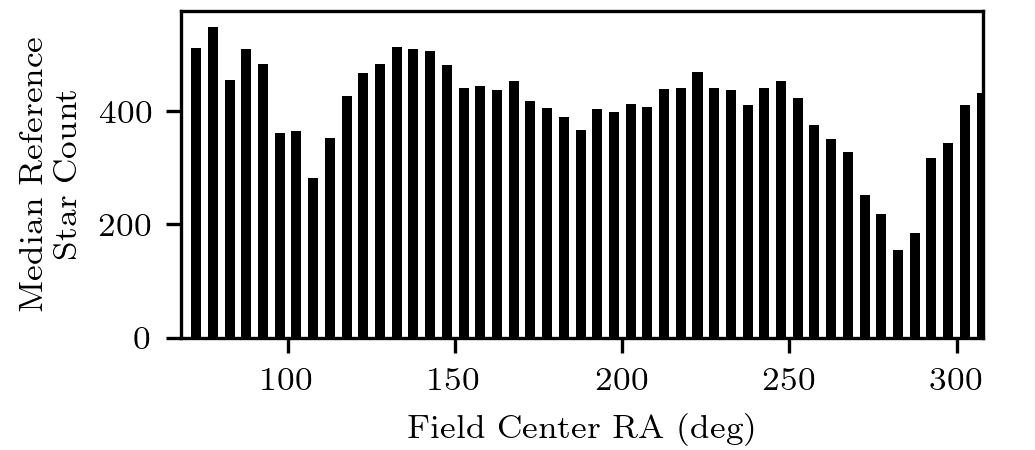}
    \caption{The median number of reference stars available for WTT photometric calibration is shown as a function of RA. About 400 reference stars are typically available per image, although this drops at RA $\sim105^\circ$ and $\sim 280^\circ$, where the increased density of stars in the galactic plane reduces the number of suitably isolated comparison stars.}
    \label{fig:comparisoncount}
\end{figure}

The final check acts to filter stars that will be contaminated by neighbors when trailed over a 5 second exposure. Figure \ref{fig:comparisoncount} shows how the number of valid reference stars varies with RA; significant dips can be seen where the galactic plane crosses the field of view, where the increased source density means that most stars in the field are blended.

The GOTO observations are calibrated against the AAVSO Photometric All-Sky Survey (APASS) survey \citep{Henden2016} using the relationships shown in Table \ref{tab:goto-calibration}. The solar colors used to evaluate color-corrected zero points were obtained from \cite{Holmberg2006}.

\begin{table}[htbp]
    \centering
    \begin{tabular}{ccccc}
    GOTO & Bandpass & APASS & Color & Solar\\
    Filter &  & Filter & term & Color\\
    \hline
    B & $400-510\,$nm & B & B $-$ V & 0.64 \\
    G & $490-580\,$nm & V & B $-$ V & 0.64 \\
    R & $595-690\,$nm & r$^\prime$ & g$^\prime$ $-$ r$^\prime$ & 0.45 \\
    \end{tabular}
    \caption{Relationships for the color calibration of the three GOTO color filters.}
    \label{tab:goto-calibration}
\end{table}

For both the WTT and GOTO data sets, catalog entries are associated with detected sources using the standard Astropy catalog matching functions. A 2D WCS distortion polynomial and the zero point color correction are fitted simultaneously with a 3-sigma outlier rejection to minimize the position and brightness residuals.

This dual-rejection is key to obtaining a reliable WCS fit in the corners of the images, preventing the fit from becoming stuck in a local minimum. The first iteration of the fit may reject all the stars in the corners of an image, but a successful fit to the center of the image will produce an improved distortion model that allows more stars to be correctly matched during the next iteration. The constrained fit thus grows outwards until the solution converges, covering the full image. This technique produces a significantly more reliable fit than Astrometry.net alone, and has now also been implemented in the NGTS forced-photometry pipeline (Gill et al, in prep).

A final improved zero point measurement is calculated by replacing the automatically extracted flux measurements provided by the source detection with aperture photometry measurements integrated over rectangular apertures covering the reference star streaks.

The WCS refitting and robust reference star selection are key to maximizing both the precision and the accuracy of the per-image zero point measurements. These metadata are saved alongside the raw frame data for use in subsequent steps of the pipeline.

\subsection{Object Detection}
\label{sec:reduction-detection}

The two-line element (TLE) history for the six RSOs of interest were downloaded from the Space Track \citep{WebSpaceTrack} service, and the TLE with the closest epoch to each night's observation was used as the basis for source detection.

\begin{figure}[htbp]
    \centering
    \includegraphics{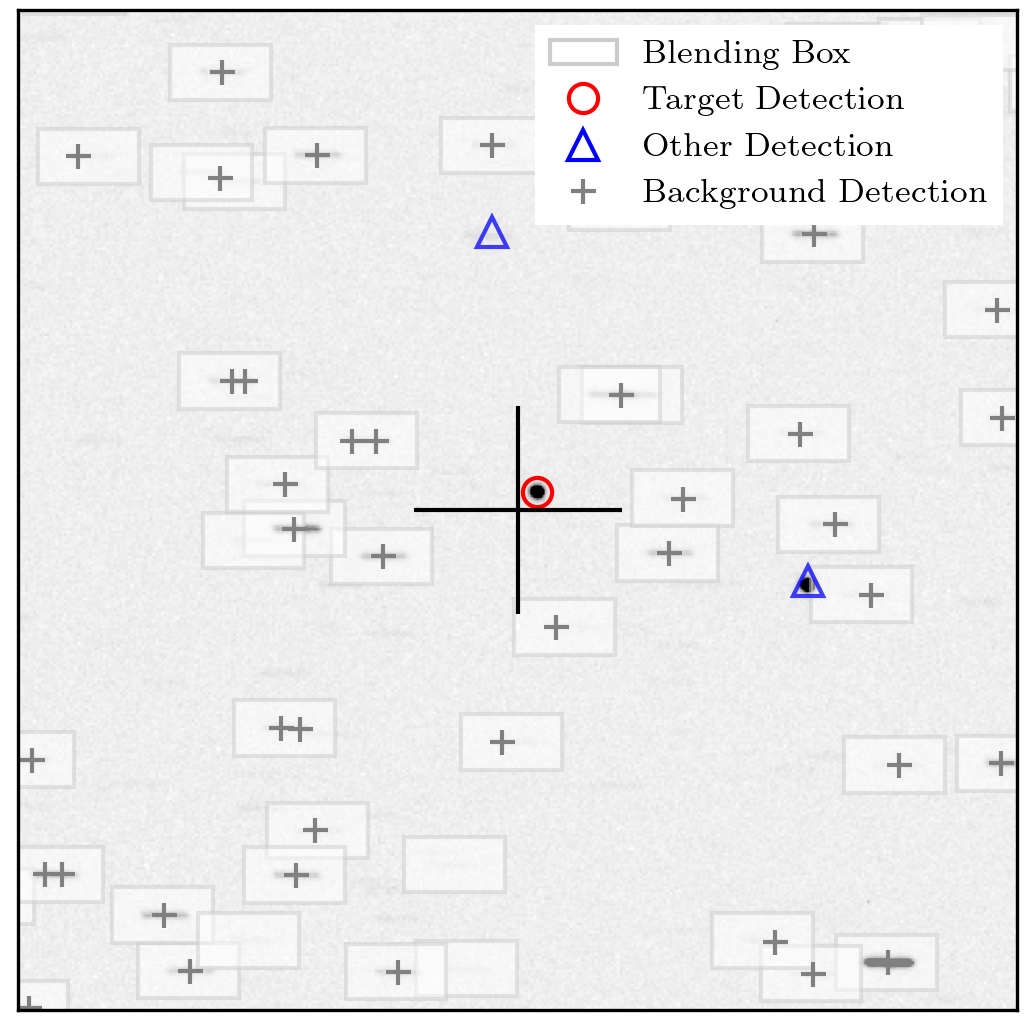}
    \caption{A WTT image cut-out demonstrating the first step of the TLE source detection procedure. For each target of interest, sources are detected in a few-arcminute window centered on the TLE position. Sources that lie within the blending box of a background star are discarded, and the remaining source that is closest to the TLE position is recorded as a candidate for the target position.}
    \label{fig:satellitedetection}
\end{figure}

The TLEs were evaluated at each exposure mid-time to find the expected position of each target in the image. Point-source detections were made within a few-arcminute sized window around each position: typically $3'$, but a larger $5'$ for MEV-2 and smaller $1.3'$ for IS10-02 to account for the large inaccuracy of the MEV-2 TLEs when it was maneuvering. Sources that coincided with expected star streak rectangles were rejected as shown in Figure \ref{fig:satellitedetection}, and the closest remaining source to the expected TLE position was recorded as a candidate detection.

\begin{figure}[htbp]
    \centering
    \includegraphics{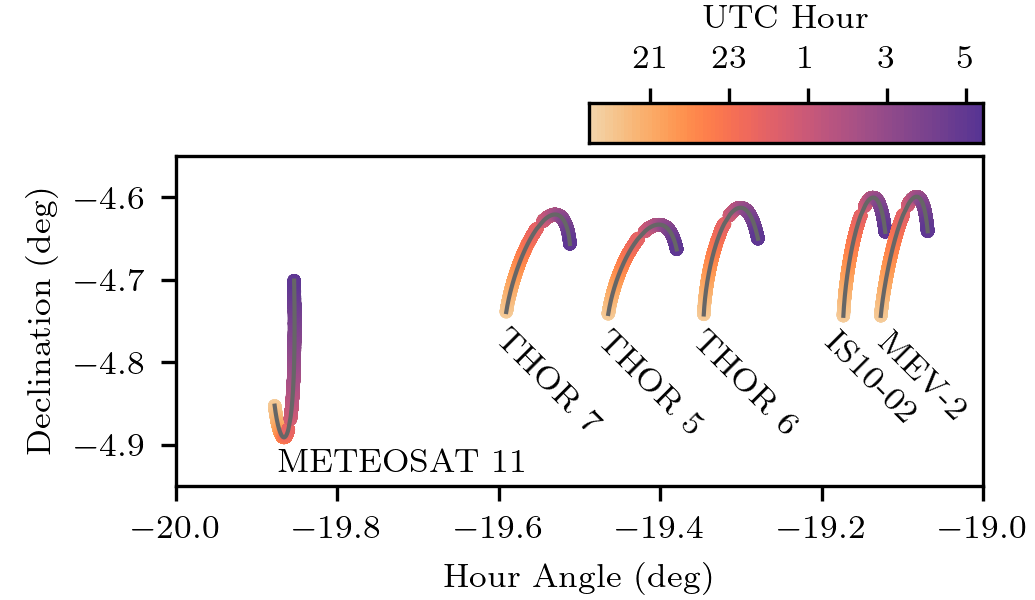}
    \caption{Candidate WTT detections for the 6 targets of interest are plotted over the course of an observing night and overlaid with the polynomial fit in gray.}
    \label{fig:trackfitting}
\end{figure}

The candidate detections for each object over the night were fitted with a smooth polynomial as demonstrated in Figure \ref{fig:trackfitting}. This reduces frame-to-frame jitter and spans the gaps where detections were missed due to e.g. star streaks overlapping the target position. $3\sigma$ outliers were rejected to exclude mis-tagged detections.

The timestamps in the GOTO observations contained uncertainties at the 1 -- 2 second level, which combined with positional jumps due to wind-shake make it challenging to model a smooth track across multiple observations. For these data, the detected positions were used directly and a small number of mis-tagged detections were removed manually.

\begin{figure}[htbp]
    \centering
    \includegraphics{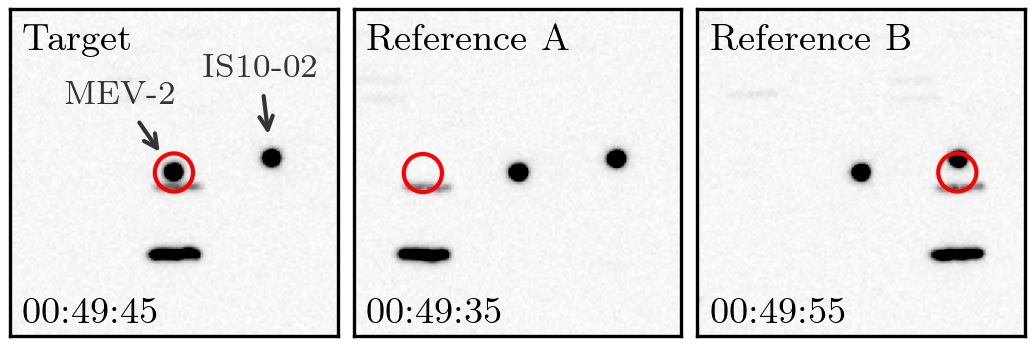}
    \caption{WTT Image cut-outs demonstrating the photometric extraction procedure. A circular aperture is placed at the target position in the target frame, and at the same RA/Dec coordinates in two selected reference images to measure the background star contamination. The minimum reference flux is selected from reference frames earlier in time (stars to the left of the target) and later in time (stars to the right) to provide a degree of robustness against nearby RSOs contaminating the reference flux measurement.}
    \label{fig:wcsresiduals}
\end{figure}

\subsection{Light Curve Extraction}
\label{sec:reduction-extraction}

Light curves were extracted by integrating circular apertures with several radii between 4 and 10 pixels at the positions evaluated from the track polynomial for each target. The aperture radius was selected for each target based on manual inspection of the light curves in order to minimize photometric scatter - smaller apertures were needed for fainter targets to reduce the impact of sky/read noise, while a larger aperture was needed for the brighter IS10-02 to prevent seeing/focus fluctuations from introducing systematic deviations.

In principle, selecting different aperture sizes for targets versus the background star streaks may introduce a systematic brightness offset due to flux lost outside the aperture (which will drift with focus changes and seeing); in practice (by investigating the differences in brightness between the smallest and largest apertures) this was found to be sufficiently small ($\leq 0.05$ mag) to neglect in this study for the sake of computational efficiency. A full treatment of this correction would involve calculating image zero points for each aperture size used by the targets.

Instrumental fluxes were converted to a magnitude using the zero point for each image. Measurements contaminated by background stars streaking through the aperture were identified by placing an apertures at the same RA, Dec coordinate in another image, providing a measurement of the background flux where the streaks are offset and isolated from the target. 

Figure \ref{fig:wcsresiduals} demonstrates that the measurement offset in one direction may itself be contaminated by the flux of a different RSO, so the minimum value from measurements on both the left and right of the target was taken as representative of the blended flux. Target measurements that were contaminated by more than 5\% blended flux were discarded.

Photometric uncertainties were calculated using the standard CCD equation formalism \citep[e.g.][]{Howell2006} that accounts for the read and photon shot noise (dark current is negligible for these short exposure times).

The rolling-shutter readout of the Marana camera creates a position-dependent timing bias. It takes $20.52\,\mu$s to digitize a single row on the sensor, leading to a time skew of $42\,$ms between the first and last rows on the sensor. This is corrected by adding to each measurement's timestamp an offset of $20.52\,\mu$s multiplied by the $y$ coordinate of the aperture position.

\section{Results}
\label{sec:results}

\subsection{Positional Accuracy}

The SAFRAN passive RF data provide an independent and significantly more precise measurement of RSO positions and velocities than can be achieved with our optical systems. This allows it to be used as a ``ground truth'' reference against which to compare the accuracy of the WTT images and reduction pipeline.

The SAFRAN CCSDS OEM positions were converted to an apparent RA and Dec using Skyfield, then interpolated to the exposure mid-times of each image.

The timestamps in the WTT images record the NTP-synchronized PC clock at the time that the camera driver returned the image data. This is delayed by a small fraction of a second from the true exposure end time due to USB transfer and driver processing delays, in addition to any error (positive or negative) in the computer clock.

The astrometric solution in each image is measured relative to the stars, which appear to move east-west at a rate of 15 arcseconds per second. A small offset in the recorded timestamps can therefore introduce a systematic offset of several arcseconds in the measured RSO positions (1 arcsec $\approx180\,$m physical distance at GEO distances). This offset affects all objects in the frame equally (after accounting for the rolling-shutter offset of up to 42\,ms) and may drift slowly during an exposure sequence as small clock rate errors accumulate. This makes it possible to qualitatively separate timing induced errors from other types of positional error. It is clear, however, that hardware-level time-stamping is crucial for any quantitative studies of object positions.

\begin{figure}
    \centering
    \includegraphics{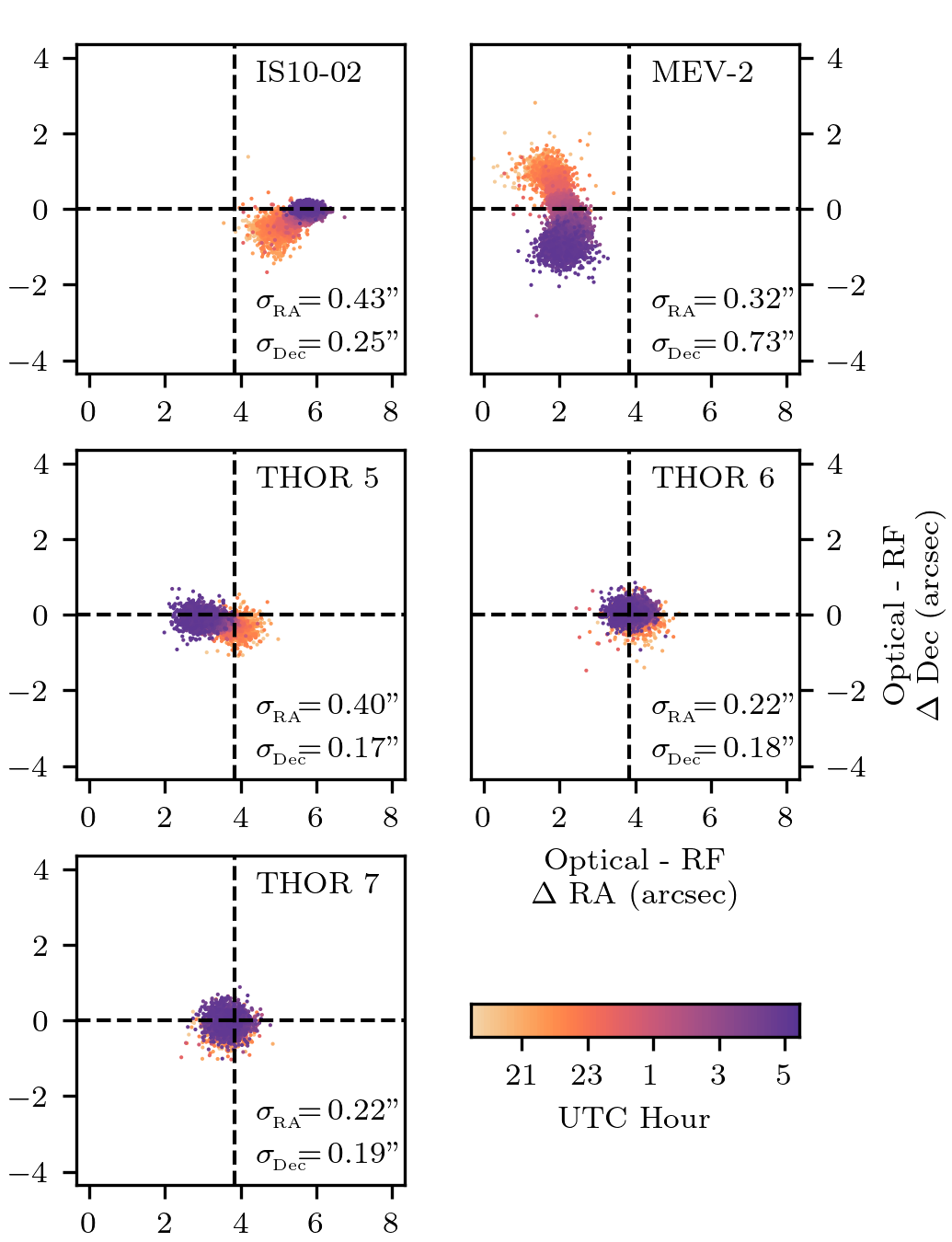}
    \caption{Centroid positions from the WTT images are compared with the ``truth'' positions calculated from the SAFRAN state vectors for the night of 2021 March 3. Each plot is presented using a $2\times2$ grid (dashed) representing the $4.5^{\prime\prime}$ WTT pixel size. A systematic RA offset of $\simeq\,$3.9 arcsec suggests that the NTP timestamps on the images were delayed by $\simeq\,$0.26 seconds compared to the true exposure epochs. The residual systematic offsets visible for MEV-2, in particular, may be caused by errors in the SAFRAN state vectors introduced by unaccounted for maneuvers.}
    \label{fig:safran-comparison}
\end{figure}

Figure \ref{fig:safran-comparison} demonstrates this comparison for one night where the timing offset remained stable. We see that the positions generally agree to within a fraction of a WTT pixel, giving us confidence in the behavior of the astrometric calibration. IS10-02, MEV-2, and to a lesser extent Thor 5 show larger systematic offsets that change with time: this may be due to errors in the SAFRAN ephemerides introduced by their orbit model failing to detect and account for small orbital maneuvers (B. Guillot, private comm).

\subsection{Phase-angle Light Curves}

With such a large volume of data, it can be particularly insightful to display the light curves as a two-dimensional map, where the equatorial phase angle varies along the horizontal axis, and date along the vertical axis. The mean brightness within each $0.5^\circ$ phase bin is represented using color. These maps make it easier to track features that drift in phase, but comes at a cost of reducing the dynamic range available for visualizing small brightness changes.

\begin{figure*}
    \centering
    \includegraphics{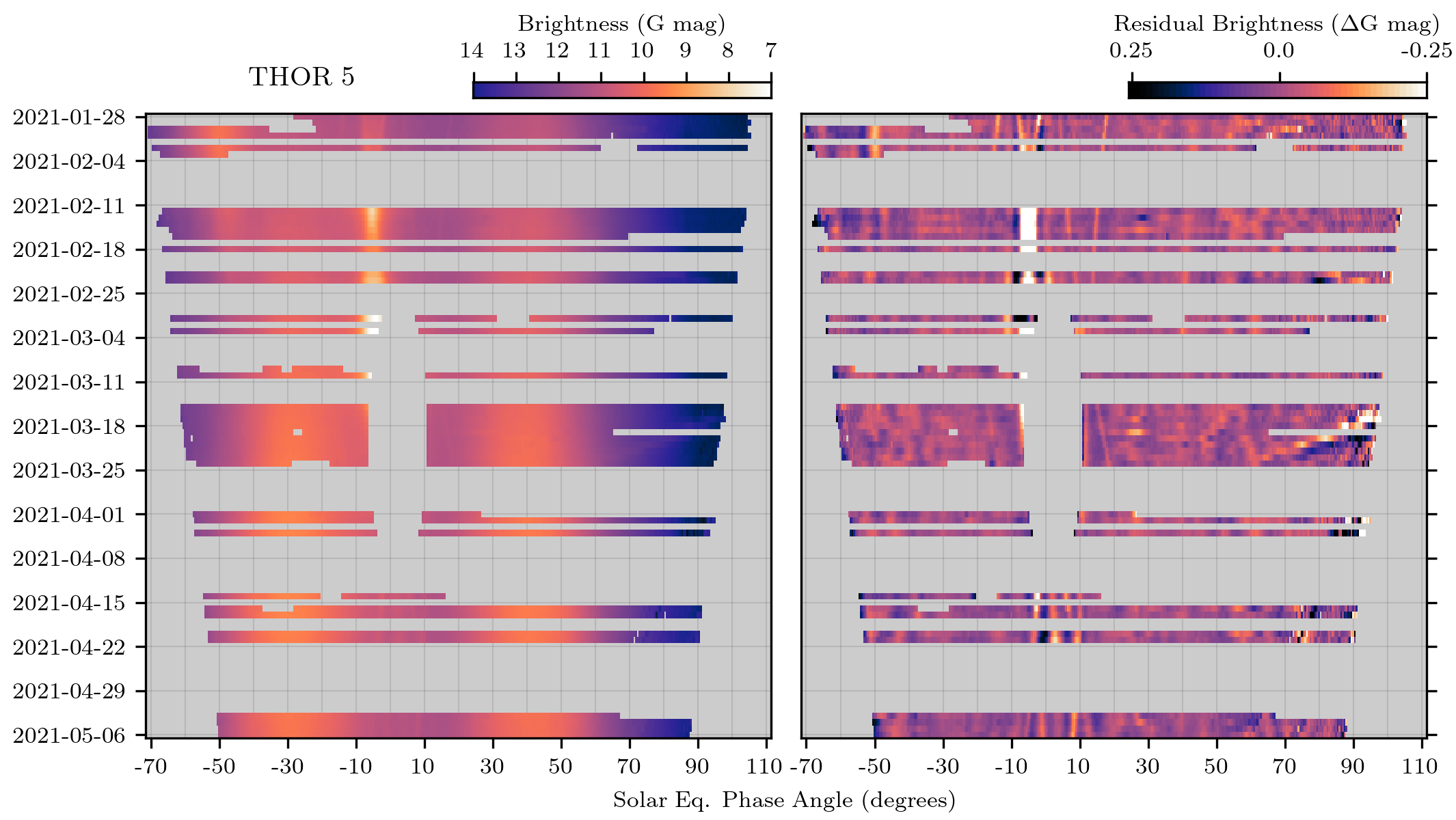}
    \caption{\textbf{Left:} Light curves of Thor 5 (SATCAT \#32487) between 2021-01-28 and 2021-05-06. Each row traces the measured intensity (calibrated against Gaia G magnitudes using the indicated color scale) across the span of a single night, plotted as solar equatorial phase angle. The vertical axis shows the evolution of the nightly brightness variations over the observation campaign, which is primarily driven by the solar declination. \textbf{Right:} Subtracting a smoothed fit of the light curve map acts as a high-pass filter to boost the contrast of short-period features, and reveals the presence of sharp, low amplitude glints that move in a consistent manner across many nights. These are likely due to specular reflections from inclined surfaces on the body of the satellite.}
    \label{fig:phasecurves-thor5}
\end{figure*}

\begin{figure*}
    \centering
    \includegraphics{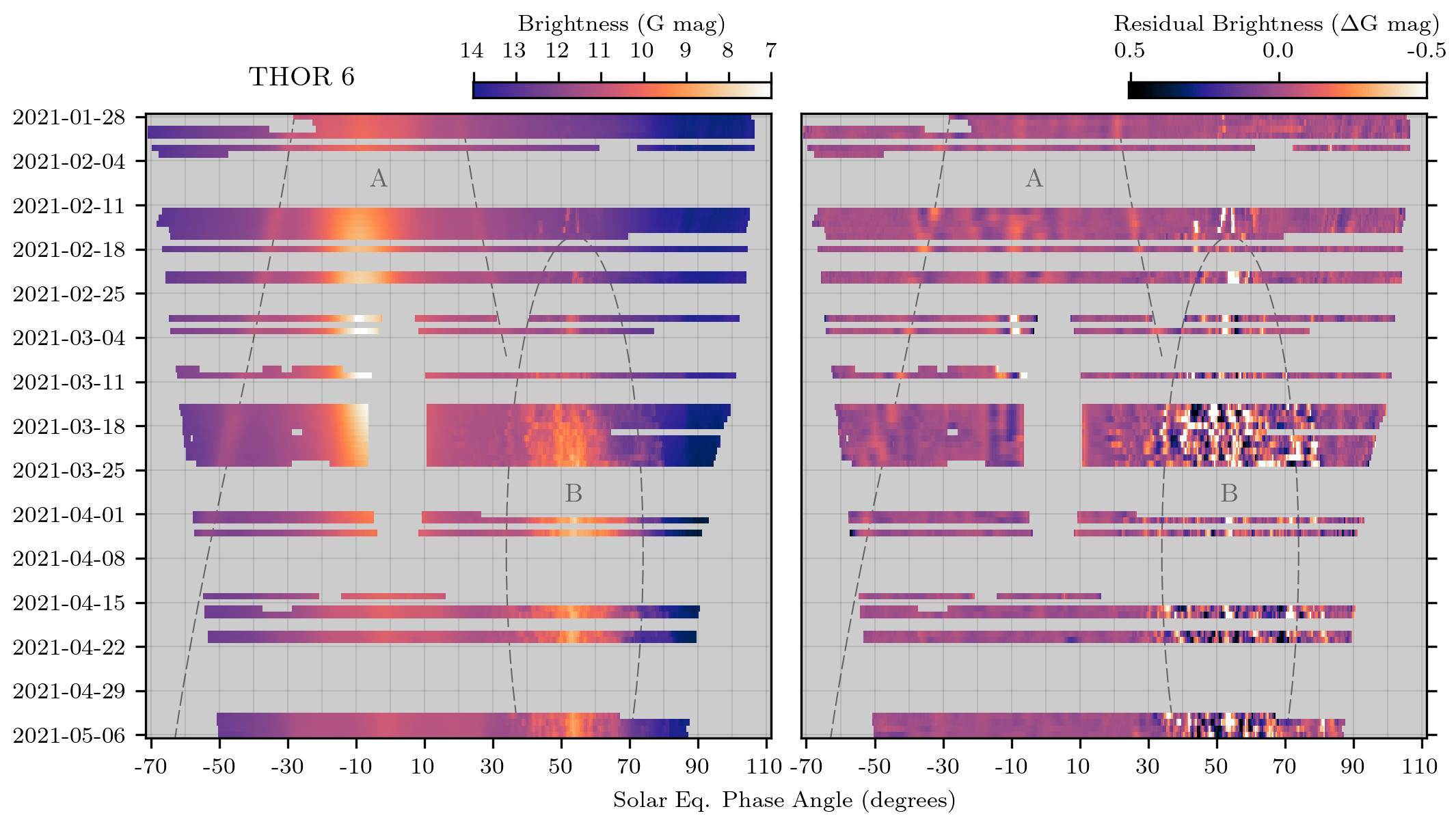}
    \caption{Light curves of Thor 6 (SATCAT \#36033) between 2021-01-28 and 2021-05-06. See Figure~\ref{fig:phasecurves-thor5} and the main text for more details. Two features (A and B) are denoted with dashed lines and explored in more detail in Section \ref{sec:thor6-analysis}.}
    \label{fig:phasecurves-thor6}
\end{figure*}

\begin{figure*}
    \centering
    \includegraphics{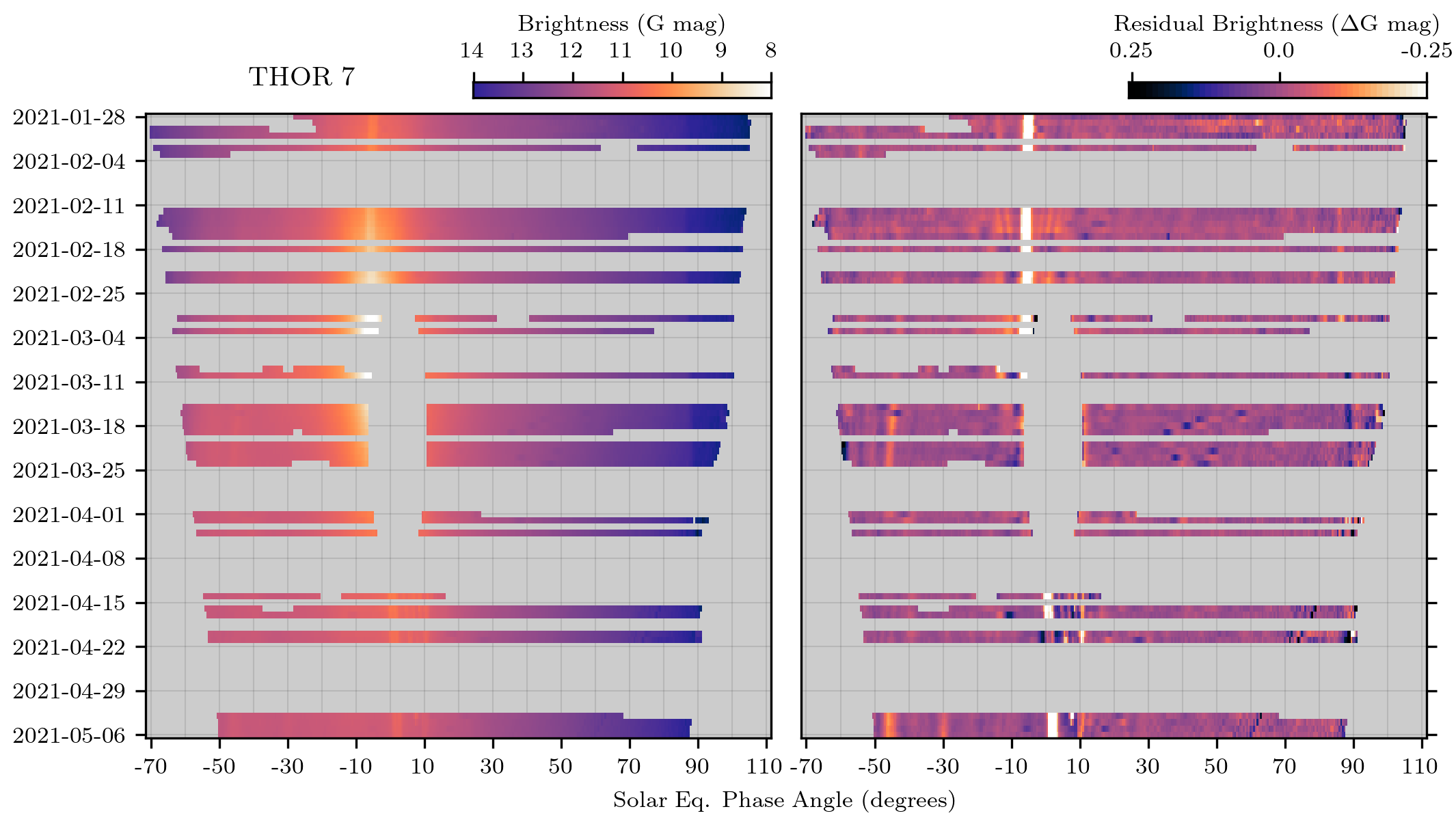}
    \caption{Light curves of Thor 7 (SATCAT \#40613) between 2021-01-28 and 2021-05-06. See Figure~\ref{fig:phasecurves-thor5} and the main text for more details.}
    \label{fig:phasecurves-thor7}
\end{figure*}

\begin{figure*}
    \centering
    \includegraphics{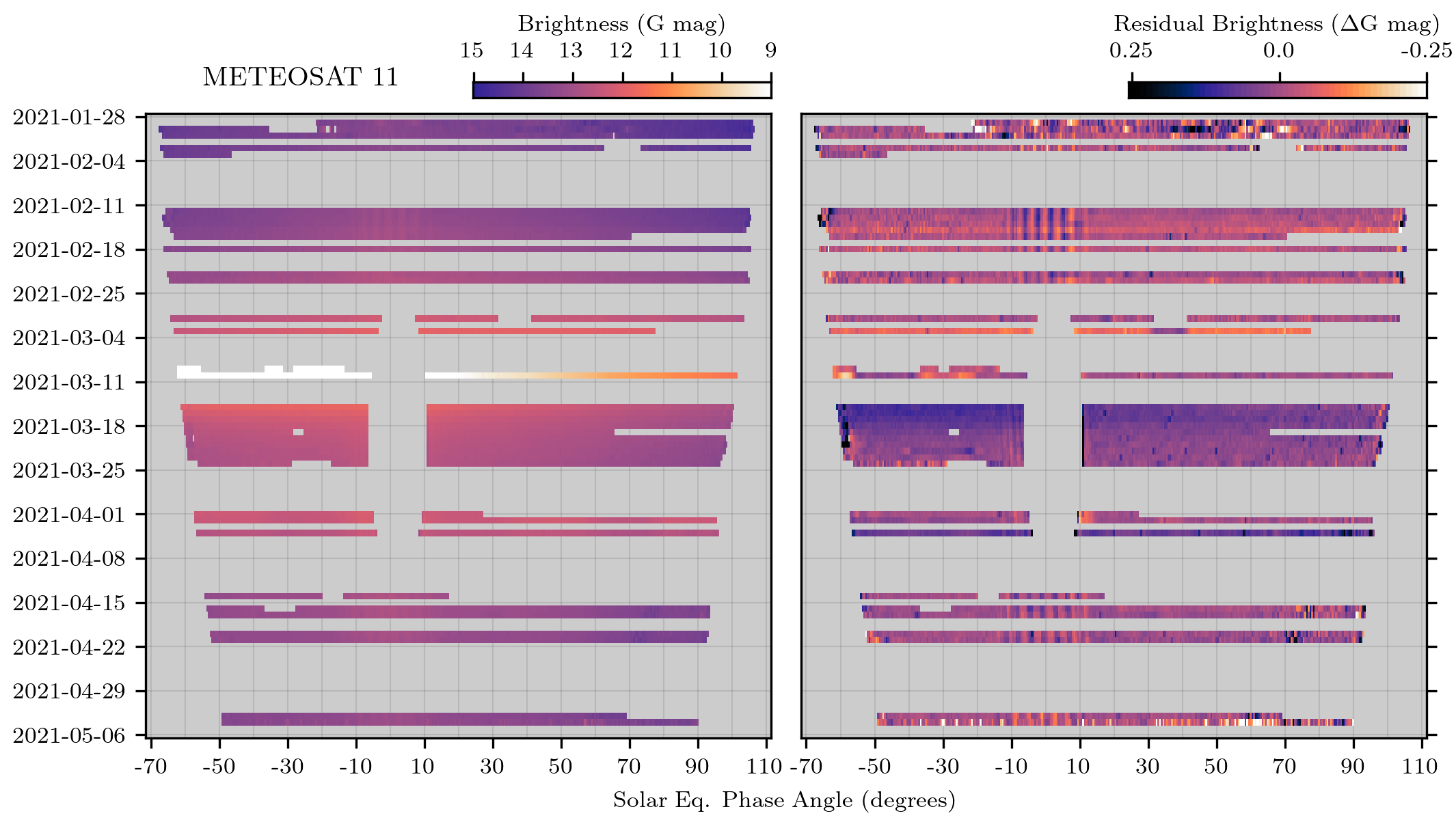}
    \caption{Phase light curves of Meteosat 11 (SATCAT \#40732) between 2021-01-28 and 2021-05-06. See Figure~\ref{fig:phasecurves-thor5} and the main text for more details.}
    \label{fig:phasecurves-meteosat11}
\end{figure*}

\begin{figure*}
    \centering
    \includegraphics{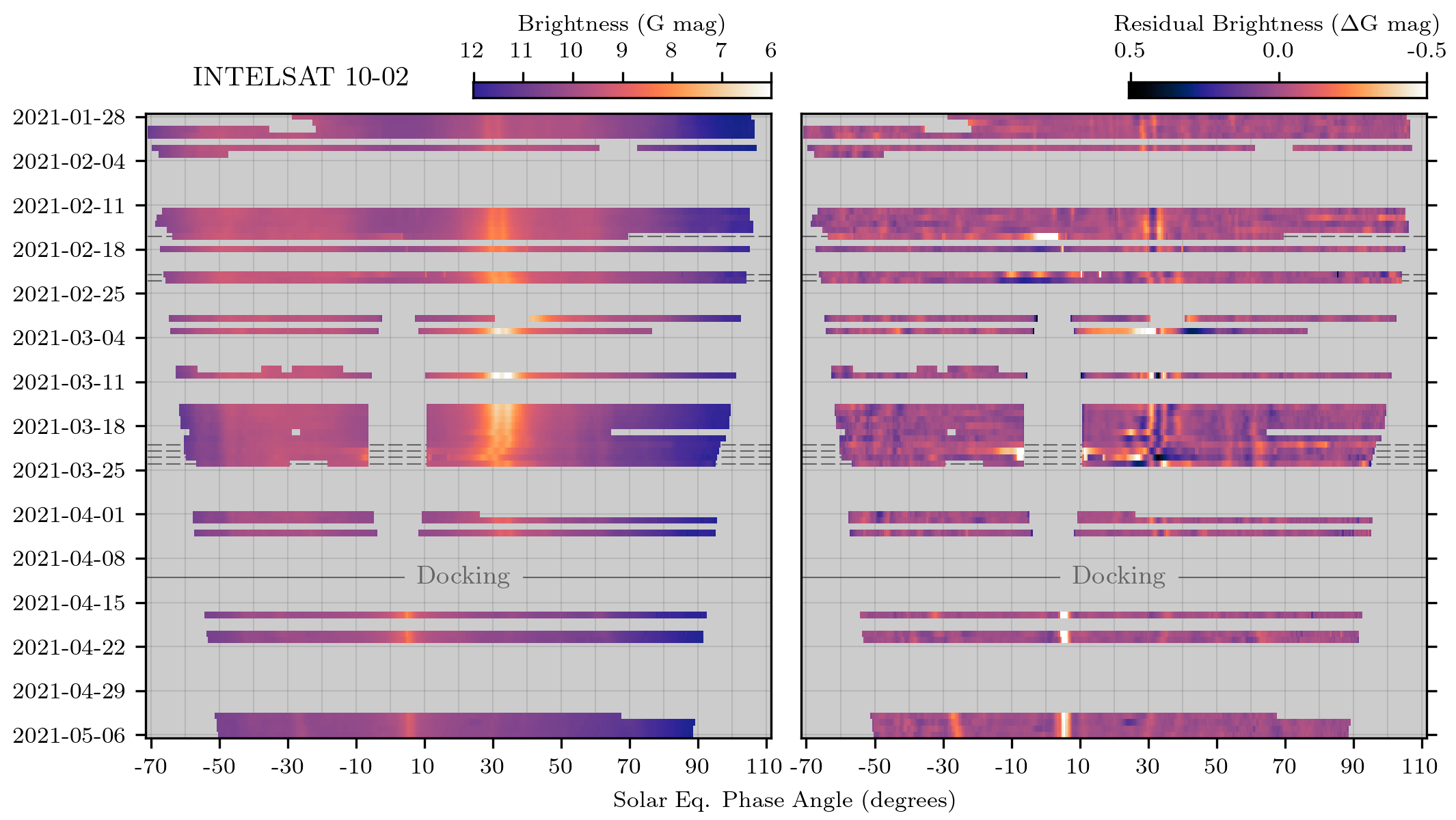}
    \caption{Phase light curves of Intelsat 10-02 (SATCAT \#28358) between 2021-01-28 and 2021-05-06. Dates where IS10-02 and MEV-2 were blended on the detector are indicated with dashed lines. See Figure~\ref{fig:phasecurves-thor5} and the main text for more details.}
    \label{fig:phasecurves-intelsat}
\end{figure*}

\begin{figure*}
    \centering
    \includegraphics{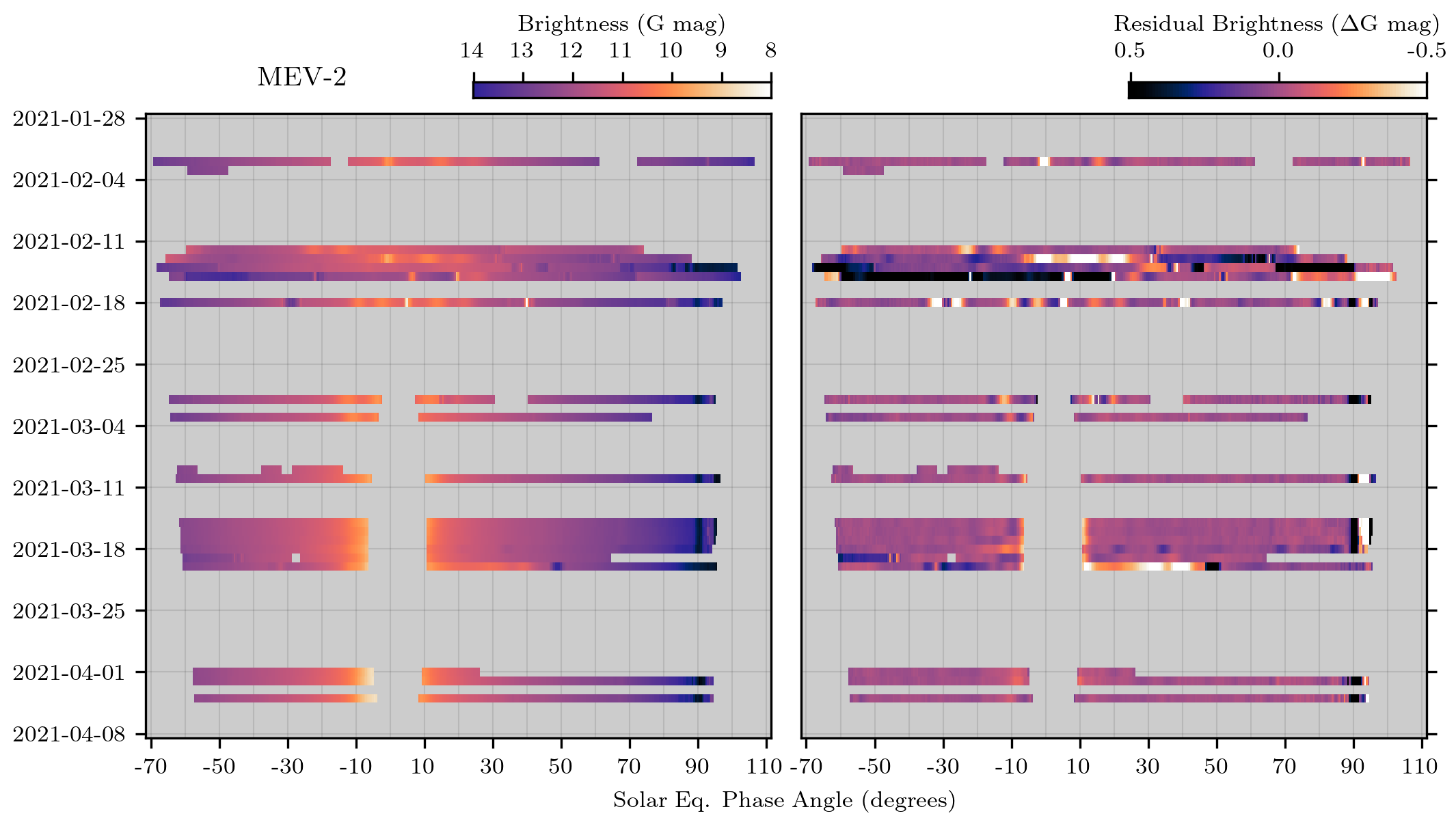}
    \caption{Phase light curves of MEV-2 (SATCAT \#46113) between 2021-01-28 and 2021-04-04, prior to its docking with IS10-02. See Figure~\ref{fig:phasecurves-thor5} and the main text for more details.}
    \label{fig:phasecurves-mev2}
\end{figure*}

Figures \ref{fig:phasecurves-thor5} -- \ref{fig:phasecurves-mev2} present light curve maps for the six RSOs monitored during the observation campaign. The left-hand panels show the large-scale brightness changes that occur within a night and evolve over the observation campaign. The right-hand panels show the residual brightness changes that remain after subtracting a smoothed version of the left-hand panels (calculated using SEP's background map functions), which reveal additional low-amplitude features that are superimposed on top of the large-scale changes. The elliptical `hole' near phase angle $0^\circ$ between March 1 - April 4 corresponds to times where the RSOs go into eclipse from the Earth's shadow.

The broad light curve features of Thor 5 (Figure \ref{fig:phasecurves-thor5}, left panel) match the `Telstar' class of \cite{Payne2006}. There is a sharp peak at $\approx-5^\circ$ phase angle, plus additional broad peaks near $\pm40^\circ$ that are believed to be associated with reflection from parabolic reflectors \citep{Wang2018}. After subtracting these features (Figure \ref{fig:phasecurves-thor5}, right panel), we see the presence of at least five short ($\lesssim 1\,$min), faint ($\approx0.1\,$mag) glints that move smoothly within $\pm30^\circ$ phase angle. These are likely associated with reflections from small features on the satellite bus; the movement in phase angle is likely to be associated with the changing solar declination angle rather than the physical movement of components, and in principle could be inverted to constrain their three-dimensional structure.

Thor 6 (Figure \ref{fig:phasecurves-thor6}) shows a broader central solar panel glint than the other RSOs, plus two smaller glints (denoted A in Figure \ref{fig:phasecurves-thor6}) that move towards larger (positive and negative) phase angles over the first two months of the observation campaign. It also shows a unique feature among these six RSOs: a complicated forest of micro-glints (named such as to distinguish these faint and short features from the larger scale solar panel glints) between phase angles $+30^\circ-+70^\circ$ (denoted B). These features are explored in Section \ref{sec:thor6-analysis} below.

Thor 7 (Figure \ref{fig:phasecurves-thor7}) broadly matches Payne's `Canonical' type during the first six weeks of observation. A central brightness peak falls off monotonically towards larger phase angles. This profile changes during the last six weeks of observations; the glint shifts to a more positive phase angle and splits into multiple peaks. It is not clear whether this change was due to the operator adjusting the solar panel offset and/or RSO attitude, or whether this was a consequence of the solar illumination changing across the equinox from slightly-above to slightly-below the local horizontal plane.

\begin{figure}
    \centering
    \includegraphics{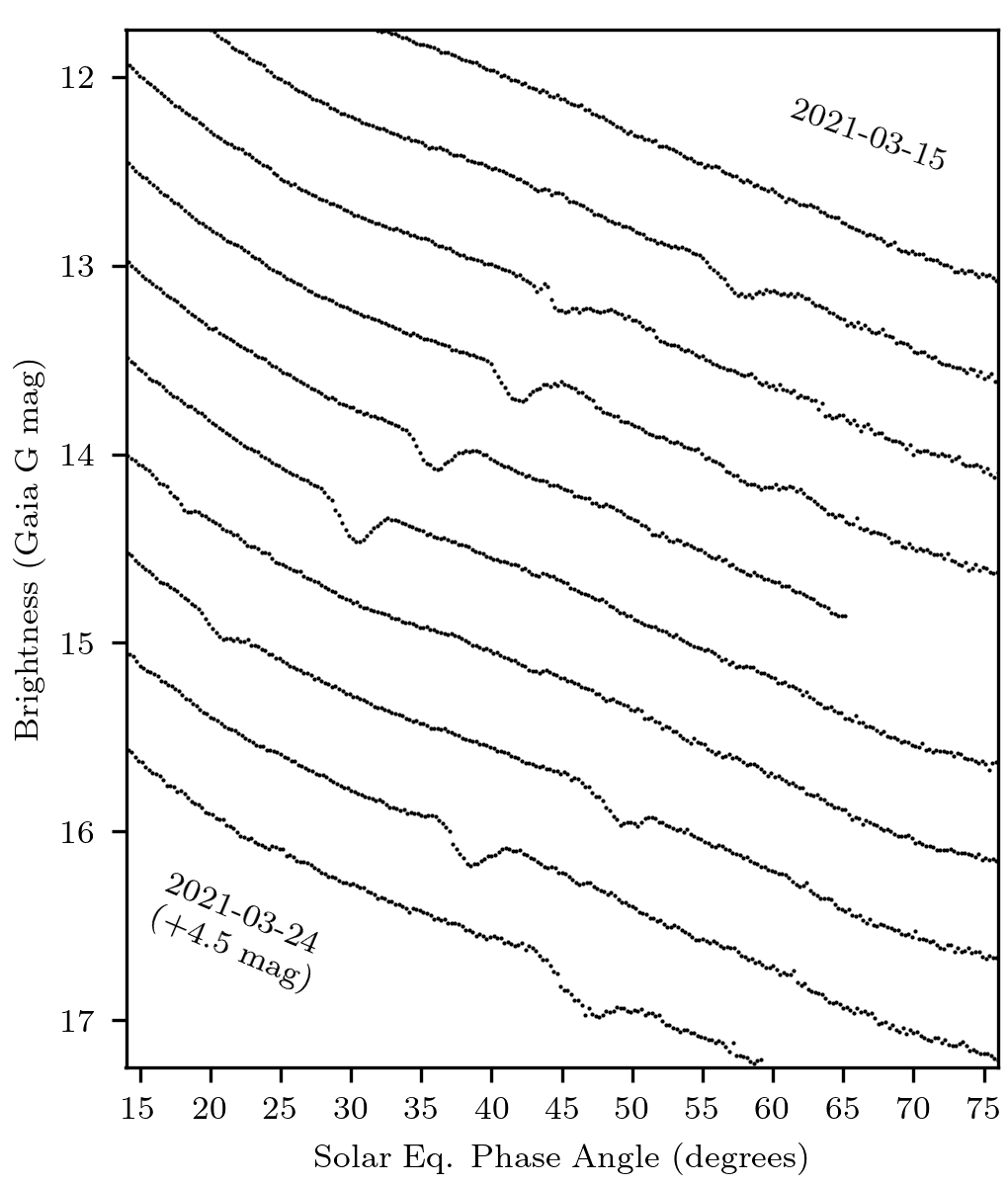}
    \caption{Light curve segments of Thor 7, binned to 1 minute cadence, are shown between 2021-03-15 and 2021-03-24. The Y axis scale is set for the first night, and successive nights are offset by +0.5 mag. A series of faint dips are visible between phase angles $+20^\circ - +60^\circ$.}
    \label{fig:thor7-dips}
\end{figure}

A series of small dips (shown in more detail by Figure \ref{fig:thor7-dips}) appear in late March, at similar phase angles to Thor 6's micro-glinting. The night-to-night position changes make their origin unclear, but they are likely due to structures on the satellite body shadowing a small amount of the flux reflected from the solar panels.

Meteosat 11 (Figure \ref{fig:phasecurves-meteosat11}) is striking for its lack of light curve features compared to the other RSOs. It generally maintains a constant brightness across the entire night, which is explained by the absence of large flat solar arrays; its solar panels instead cover its spinning cylindrical body. The residual brightness plot reveals a series of seven faint glints centered around 0$^\circ$ phase angle. A significant brightening event was seen in early March, and is investigated in Section \ref{sec:meteosat-brightening}.

Intelsat 10-02's light curves before MEV-2 docked in mid-April (Figure \ref{fig:phasecurves-intelsat}) were dominated by a double-peaked glint around $+30^\circ$ phase angle. This is the usual solar-panel associated glint peak, but appears different to the other RSOs because of (a) an unusually large solar panel offset angle, and (b) the two solar panel arrays are oriented with slightly different offsets. Changes to the offset angle are visible on 2021-03-01, 2021-03-22, and 2021-03-23 -- presumably related to the RPO activities with MEV-2.

There were several nights where the projected separation of IS10-02 and MEV-2 was too small to individually resolve the two objects on the detector. The light curve signatures on these nights were dominated by the (much brighter) flux from IS10-02, but there are several instances where glints from MEV-2 contribute significant additional flux. Section \ref{sec:glint-astrometry} investigates whether these events can constrain the relative positions of the two unresolved objects.

Following the docking, the solar panels appear to have been shifted to a more usual angle near $0^\circ$, and the glint is no longer double-peaked. This change in the apparent light curve signature is presented more clearly in Section \ref{sec:is1002-analysis}.

MEV-2 (Figure \ref{fig:phasecurves-mev2}) appeared to switch between two distinct operating modes.  In one mode, it was presumably holding position, looking like a `Canonical' RSO type; at other times it showed frequent sharp changes in brightness that must be related to attitude changes while it was maneuvered in orbit around IS10-02. These maneuvers are explored in more detail in Section \ref{sec:mev2-maneuvers}.

\subsection{Color Indices}

An initial analysis of the GOTO photometry showed that the color index measurements (i.e. the brightness \textit{difference} between two color bands) were contaminated by systematic offsets between telescopes. This was an expected effect caused by imperfect flat fielding, resulting in a spatial variation in zero point of a few hundredths of a magnitude (a few percent) across the field. Correcting these systematic offsets is essential for improving the precision of the color index measurements, and their resulting potential as characterization markers.

\begin{figure*}
    \centering
    \includegraphics{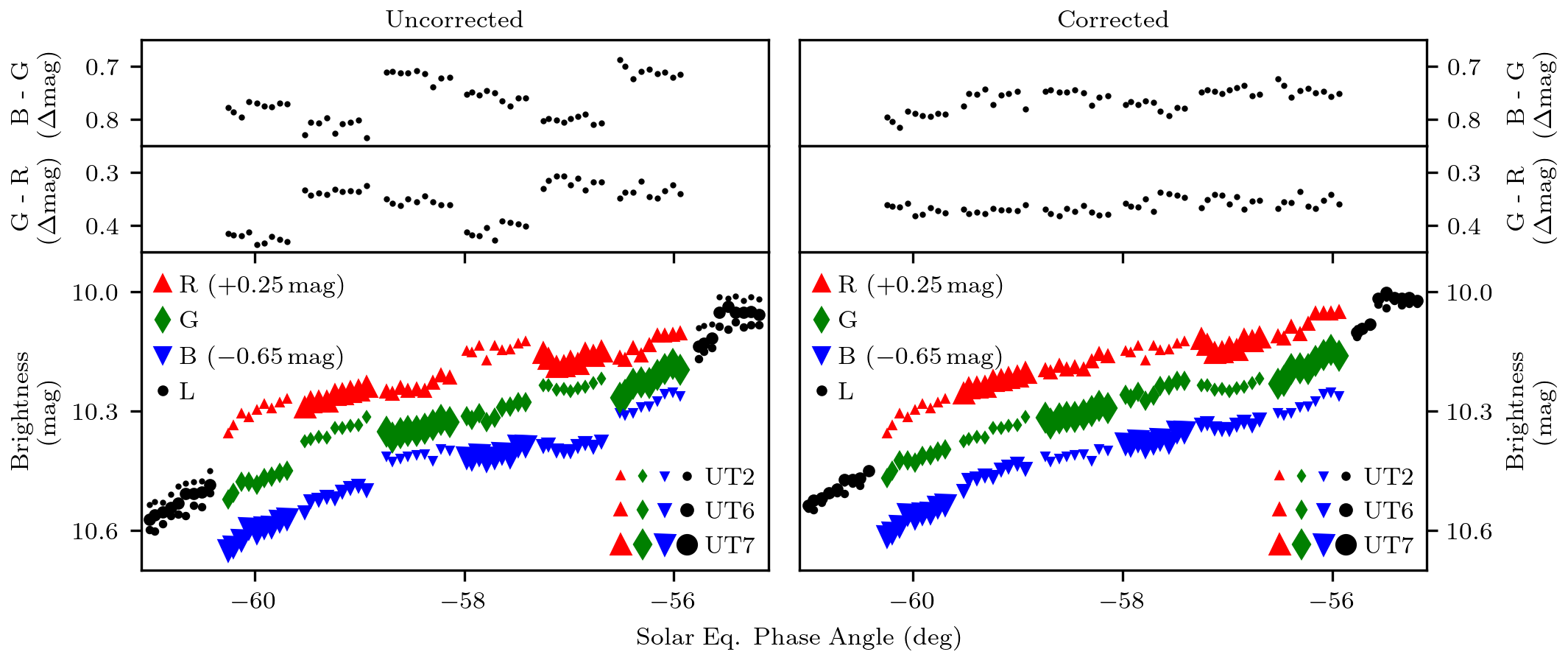}
    \caption{Color light curves of IS10-02 obtained using GOTO-North on 2021-02-02. The left panels show the as-reduced photometry, with percent-level systematic offsets between unit-telescopes. The right panels show the improved results after applying a zero point correction calculated by aligning the simultaneous L~band measurements. The R and B bands have been offset in the plots to reduce white space and accentuate the impact of these offsets. The symbol shape and color in the bottom panels is used to distinguish filter type, while the symbol size identifies which unit telescope the measurement was from.}
    \label{fig:goto-example}
\end{figure*}

These variations were found to vary with position on the detector, but not with color, which allowed them to be measured and corrected for the overlapping region by including images in the exposure sequence where the three telescopes were observing simultaneously with the same filter. The L band (broad optical) filter was chosen for these calibration images due to its wider band pass producing higher signal-to-noise measurements. The mean brightness offsets between each pair of telescopes were found for each object, and applied to correct the systematic offsets in the color observations. This procedure is demonstrated in Figure~\ref{fig:goto-example}.

The B$\,-\,$G and G$\,-\,$R color indices were typically constant across the 30 minute window of the GOTO observations. The measurements were reduced to a single pair of color indices for each night by calculating a $3\sigma$ clipped mean and standard deviation for each observation. Several instances where the color indices of Thor 6 and Meteosat 11 were \emph{not} constant are excluded, and investigated in more detail in Sections~\ref{sec:thor6-analysis} and \ref{sec:meteosat-brightening} respectively.

\begin{figure}
    \centering
    \includegraphics[width=\columnwidth]{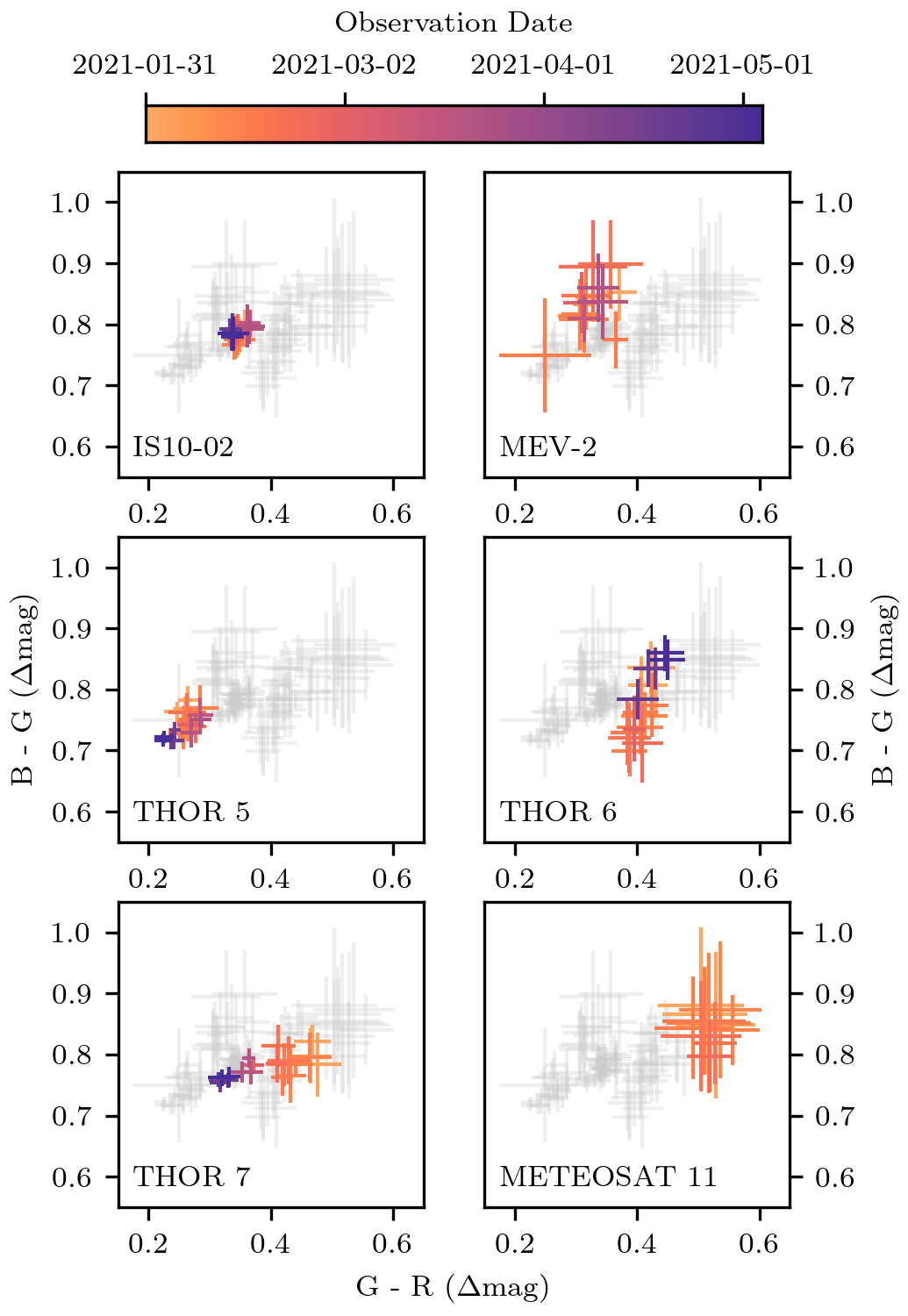}
    \caption{Color-color diagrams show that the six RSOs can be separated based on their mean colors at the start of the observing night. The three Thor RSOs show an evolution of color with time that we attribute to the changing phase angle range of the measurements introduced by the progressively later time of sunset across the observation campaign.}
    \label{fig:goto-colors}
\end{figure}

Figure \ref{fig:goto-colors} shows that the six RSOs occupy distinct areas of the color-color plane, reinforcing the idea that color indices can be used to uniquely distinguish individual objects. The color of the combined IS10-02 + MEV-2 stack is indistinguishable from IS10-02 alone.

However, we note that the three Thor satellites show evidence of color changes over the course of the observation period, which we attribute to the change in the phase angle range of our measurements: Observations in January covered phase angles $-60^\circ\pm 5^\circ$, moving to $-40^\circ\pm 5^\circ$ by May due to the progressively later sunset times. Changes in the color index as a function of phase angle complicate the prospect of using them for unique identification, but open significant opportunities for characterization when simultaneous multi-color light curves are available over a full night \citep{Airey2025}.

\section{Selected Features}\label{sec:selected-cases}

\subsection{Thor 6 Glints}\label{sec:thor6-analysis}

\begin{figure*}
    \centering
    \includegraphics[width=\textwidth]{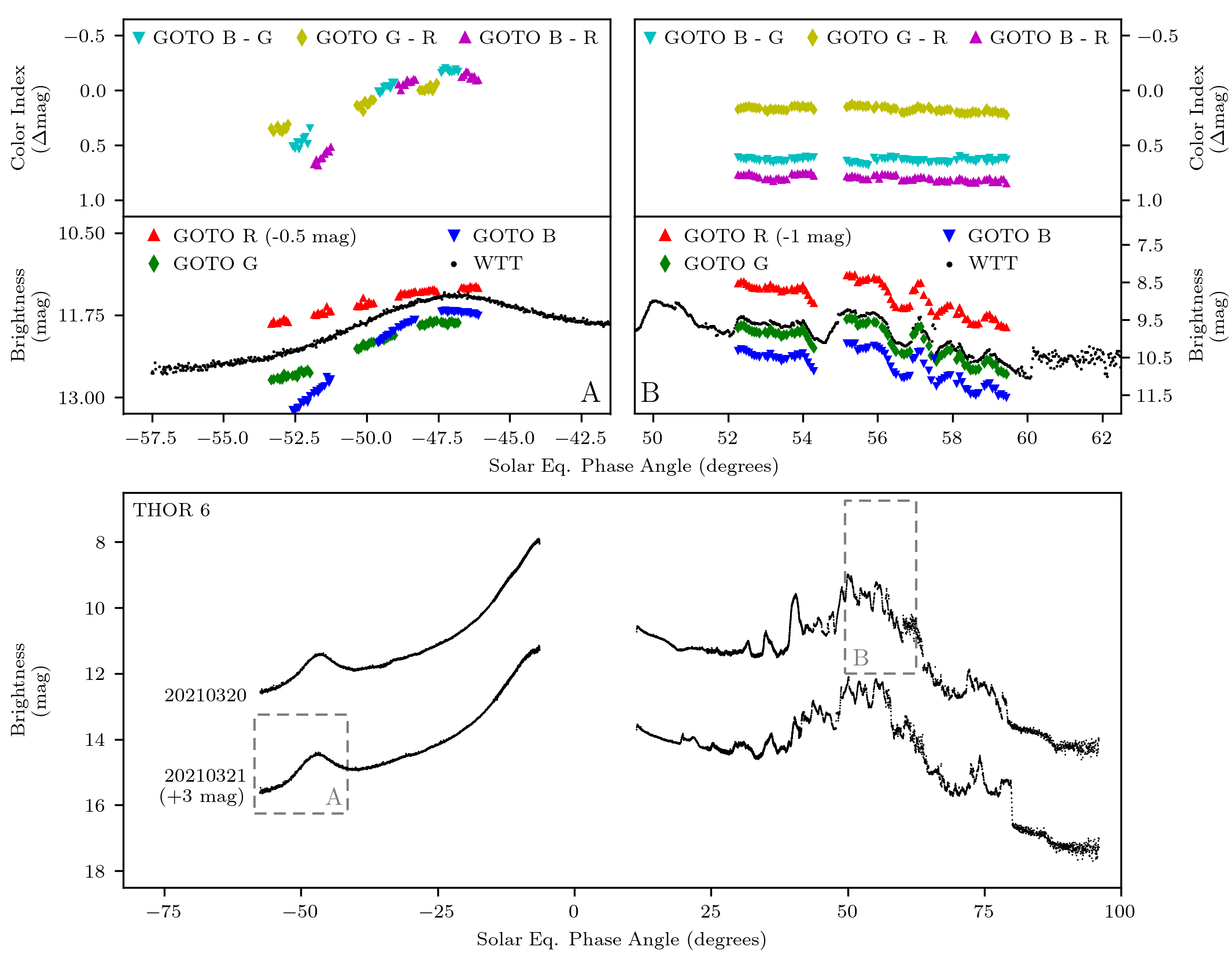}
    \caption{Simultaneous multi-color observations obtained with GOTO provide additional insight into features seen in the WTT light curves of Thor 6. Inset (A) shows that the glint features that move outward in phase are strongly blue, which suggests an origin from the solar panels. Inset (B) shows that the complicated micro-glint structures have a uniform color, which suggests that they originate from a material that has a flat spectral profile. The GOTO R band in the inset plots has been offset to improve readability.}
    \label{fig:thor6-colorpeak}
\end{figure*}

The light curves of Thor 6 showed two additional features that warranted further analysis. The first (denoted A in Figures \ref{fig:phasecurves-thor6} and \ref{fig:thor6-colorpeak}) was a pair of glints that shifted towards larger (positive and negative) equatorial phase angles over the course of the observation campaign. The position of the glints appear to track the solar declination ($\delta_\odot$), i.e. the latitudinal equivalent to the equatorial phase angle: the negative glint occurred at $-(\delta_\odot + 46^\circ)$, and the positive glint at $\delta_\odot + 39^\circ$. Color-resolved observations of this feature were serendipitously obtained when the negative peak moved into the phase range that was being monitored with GOTO. These observations, shown in Figure~\ref{fig:thor6-colorpeak}, reveal that the additional flux in this glint is strongly blue.

The second feature (denoted B in Figures \ref{fig:phasecurves-thor6} and \ref{fig:thor6-colorpeak}) is a complex series of sharp micro-glints which show a mix of features that remain stationary in phase over successive nights; that drift in phase; and that do not clearly repeat. Targeted GOTO observations (Figure~\ref{fig:thor6-colorpeak}) showed that the color remains constant during these glints, suggesting that the reflecting surfaces have a flat spectral profile.

Simultaneous high-cadence polarimetry \citep{Wiersema2022} showed that there were no clear changes in the linear polarization during these features; however individual 2-second outliers did intriguingly appear to be coincident with changes in the optical light curve.

\cite{Vananti2017} and \cite{Zilkova2023} demonstrated that many types of spacecraft material can be grouped based on the shape of their spectral profile. The strong blue color of feature A suggests that it originates from a solar cell, but the feature being offset away from the expected 0$^\circ$ phase angle would require the surface to be inclined with respect to the main solar arrays. It is common \citep[see e.g.][]{Flament1990, Lagadec2013} for box-wing RSOs to feature small `flaps' attached to their solar arrays to aid in their use as solar sails for attitude control. An important feature of these flap designs is that they are mounted with a fixed angular offset from the solar panels, and occur in pairs mounted in opposite directions. This would naturally lead to a symmetric pair of glints that appear before and after the main solar panel glint in phase angle.

The flat color signature of feature B is consistent with the silver MLI spectrum presented by \cite{Vananti2017}. Pre-launch photographs of Thor 6 \citep{WebThor6Photo} show that the outer layer of its MLI appears to be black Kapton, but this presumably also has a reflection spectra closer to flat than the red-dominated reflections from gold colored Kapton. \cite{Wiersema2022} suggest the lack of polarization changes over the glints could arise from many facets of a `wrinkly' layer of MLI washing out the polarization signal while producing many small optical glints.

\subsection{Meteosat 11 Brightening}\label{sec:meteosat-brightening}

\begin{figure}
    \centering
    \includegraphics{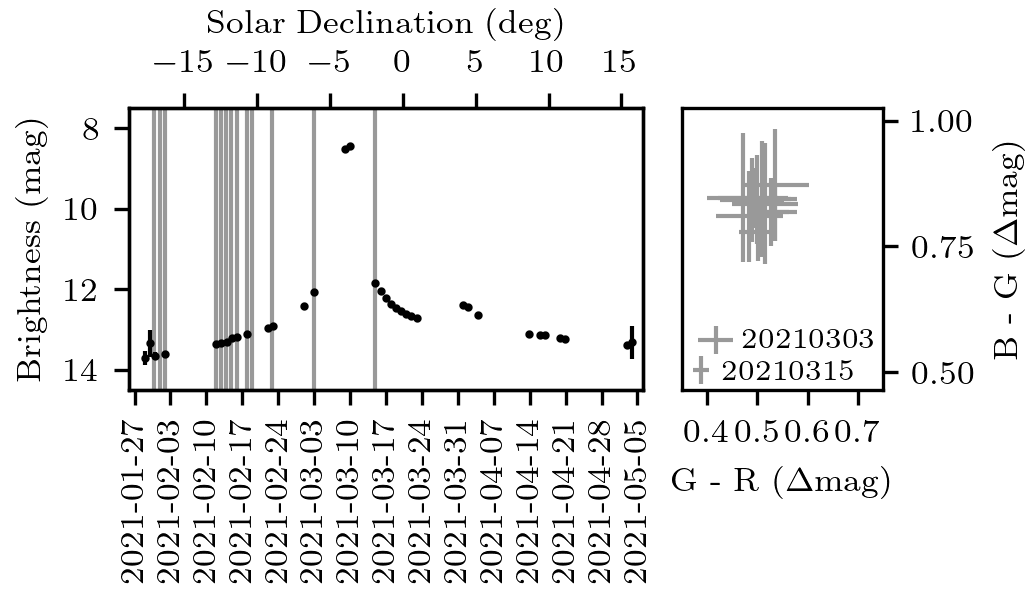}
    \caption{The mean brightness of Meteosat 11 at phase angle -20$^\circ$ was seen to spike during the days around 2021-03-10. Vertical gray lines show the dates of the GOTO color measurements, which show a clear blue-ward shift when it begins to brighten.}
    \label{fig:meteosat-glint}
\end{figure}

Meteosat 11 was observed to undergo a significant brightening event mid way through the observation campaign, reaching \mbox{G$~\approx~8.5$} on the 9\textsuperscript{th} and 10\textsuperscript{th} of March compared to its more usual \mbox{G$~\approx~12-13$}. Figure \ref{fig:meteosat-glint} shows the RSO becoming brighter, slowly at first, and bluer in the days before the event, before showing a very large increase centered on March 10\textsuperscript{th}.

Meteosat 11 has a cylindrical bus that is covered with solar cells. It spins on its cylindrical axis, which is aligned to within a few degrees of the Earth's rotation to allow its instruments to scan in rows across the Earth's disk. The observed brightening is consistent with what we would expect for a specular glint from this surface.

\begin{figure}
    \centering
    \includegraphics{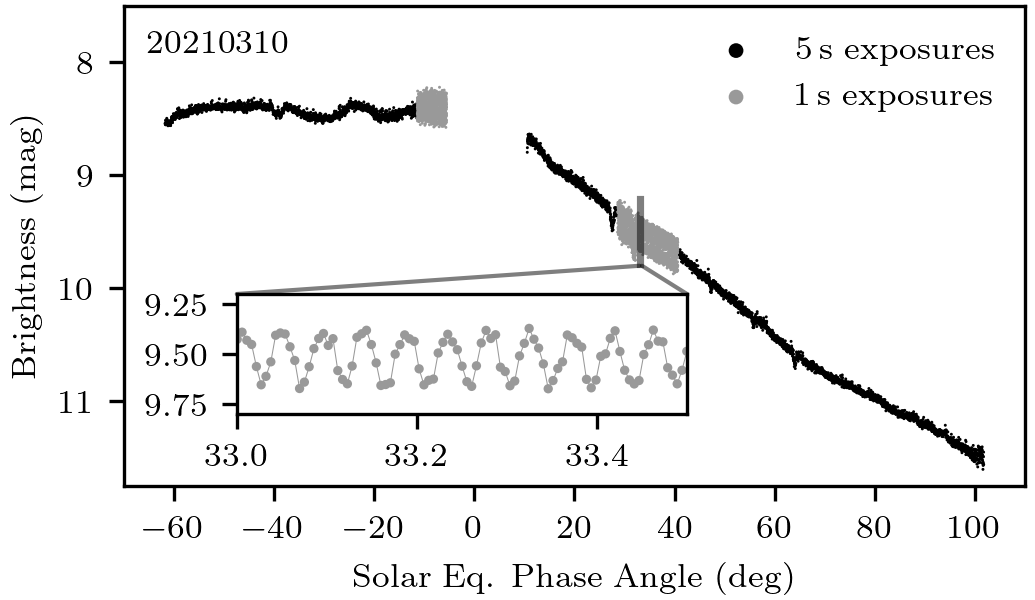}
    \caption{High cadence observations (timed to avoid saturation during the solar panel glints of the other RSOs) of Meteosat 11 during its peak brightness revealed the presence of a $11\,$s periodicity. This period matches the expected Nyquist alias period for the 100 RPM (0.6\,s) spin period.}\label{fig:meteosat-zoom}
\end{figure}

\begin{figure}
    \centering
    \includegraphics{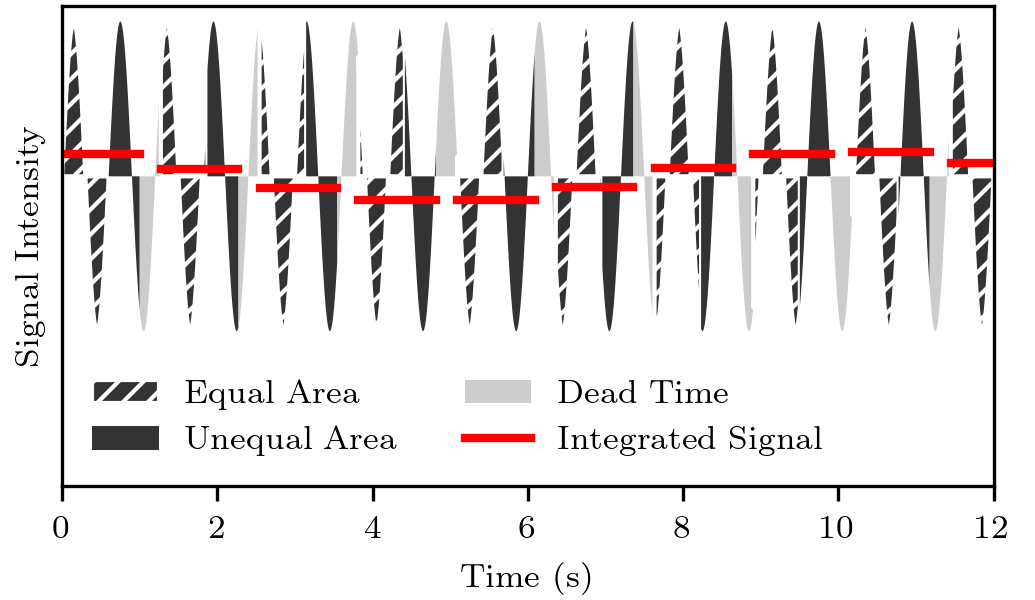}
    \caption{A visualization illustrating under-sampling a $0.6\,$s sinusoidal modulation (shaded) to produce an apparent alias period of $\approx11\,$s. The fraction of the exposure period that contains an integer number of wavelengths (i.e. $0.6\,$s) contains the same area above and below the signal mean, so does not change between exposures. The remaining fraction contains unequal areas above and below the mean, which changes from exposure to exposure due to the mismatch between exposure cadence (which may include additional dead time) and the true period. This produces a beat signal with a longer period and reduced amplitude compared to the true signal.
}\label{fig:meteosat-alias}
\end{figure}

An additional unexpected feature was seen during the high-cadence (1\,s exposure) observations on March 10th: an apparent $\approx 11\,$s sinusoidal oscillation with a peak-peak amplitude of $\approx0.25$ mag, shown in Figure \ref{fig:meteosat-zoom}. We do not believe this to be a real signal, but rather a Nyquist alias of the 0.6\,s spin period being under-sampled by the $\sim1.27\,$s exposure cadence (this is greater than the exposure time due to processing overheads on the data acquisition computer). A visual sketch illustrating this aliasing is shown in Figure \ref{fig:meteosat-alias}. Observations with a cadence shorter than 0.3\,s would be required to unambiguously confirm this hypothesis.

\subsection{Astrometric Localization of Blended RSOs}\label{sec:glint-astrometry}

One of the challenges faced by astronomical surveys for transiting exoplanets is confirming that a small periodic dip in an observed light curve of a star is caused by a planet-sized body transiting a bright star, and not by a large periodic dip from a background eclipsing binary that is diluted by a bright foreground star. The direct approach to break this degeneracy is to obtain high angular resolution imaging, but this requires additional observations using specialized instrumentation which may not be available.

An alternative technique \citep{Gunther2017} is able to constrain background blends directly from the survey images: when the point spread function (PSF) of a source is actually made up of two spatially unresolved sources, the apparent center of flux of the combined profile will shift by a small ($\approx$ milli-pixel) amount as the relative contribution of the two PSFs change with the variable stars brightness.

The same technique can, in principle, be applied to constrain the relative positions of unresolved RSOs. If one of the RSOs glints during an observation (which is likely if it is maneuvering), the center of flux will shift slightly in the direction of that object.

\begin{figure}
    \centering
    \includegraphics{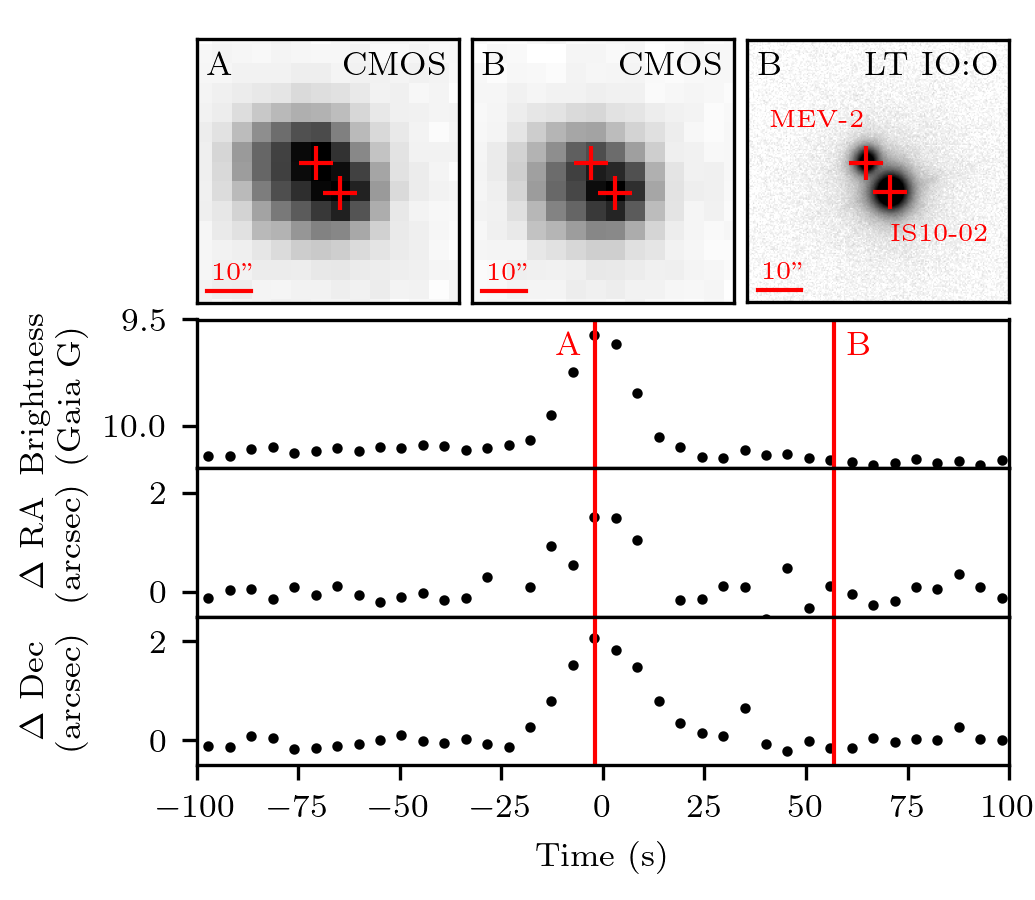}
    \caption{The flux centroid of the unresolved MEV-2 + IS10-02 point source is observed to shift during glint events (A). This offset can be used to constrain the separation of the two RSOs, as confirmed by near-simultaneous high-resolution imaging (B). Reproduced from \cite{Meredith2023}.}
    \label{fig:glintoffset}
\end{figure}

Figure \ref{fig:glintoffset} presents the centroid offset that was measured for one such event where there was coincidental near-simultaneous imaging by the higher resolution Liverpool Telescope \citep[see][]{George2021}. A detailed study of this and several other glint events has been presented in \cite{Meredith2023}, which demonstrates that the position angle of the two sources can be accurately measured, but that their separation is more challenging to constrain.

\subsection{IS10-02 Solar Panel Glints}\label{sec:is1002-analysis}

\begin{figure*}
    \centering
    \includegraphics[width=\textwidth]{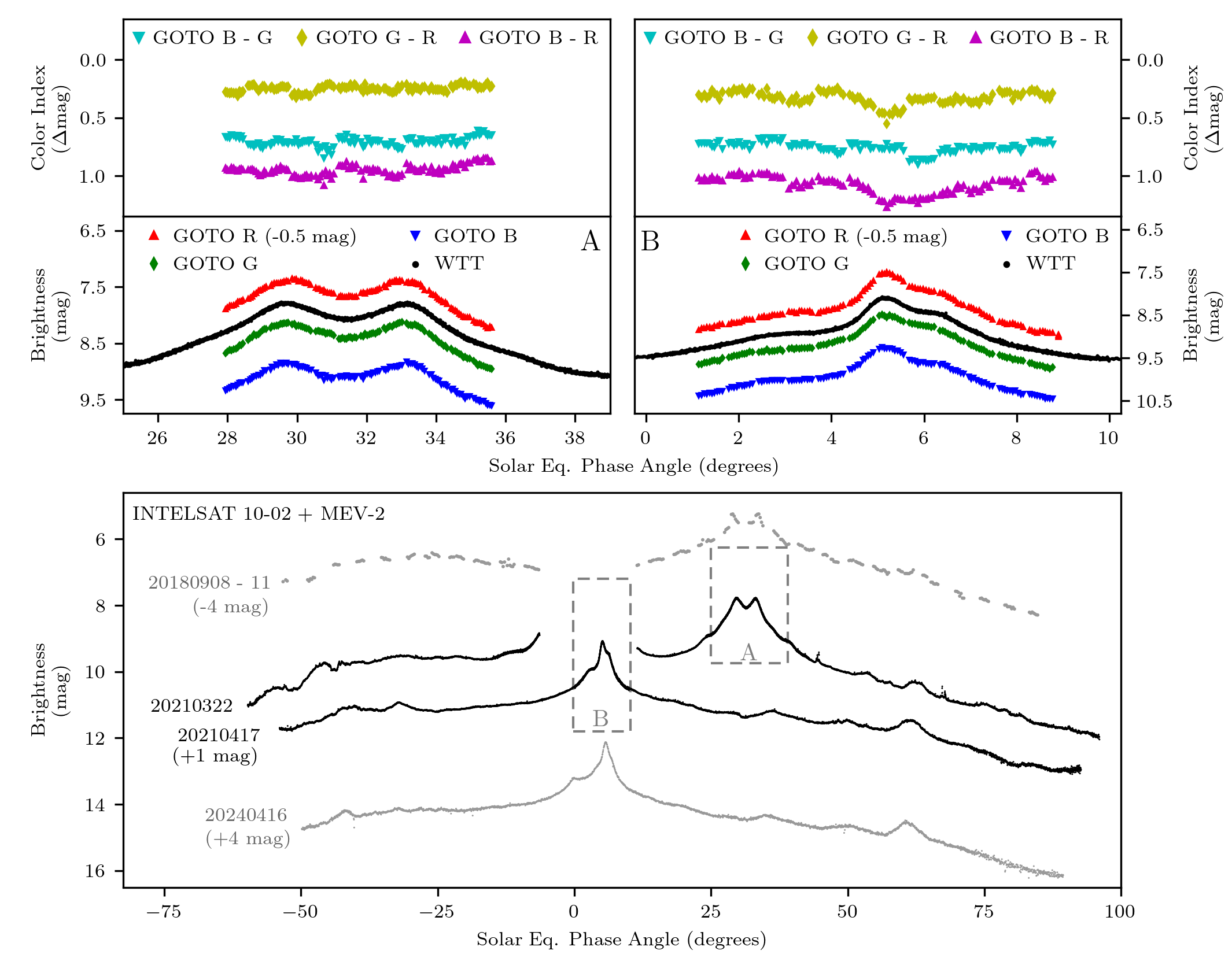}
    \caption{\textbf{Top:~}Simultaneous multi-color observations obtained with GOTO showed no significant color change across the duration of IS10-02's main solar panel glint. After docking, the data suggests a small redward shift during the brightest central peak. \textbf{Bottom:~}Full night light curves before and after docking show a total change in light curve signature, dominated by a shift in the main solar panel glint peak from $30^\circ$ to $\sim5^\circ$. Additional observations obtained $\sim 3$ years before and after docking show that these light curves are representative of IS10-02's optical signature over long timescales.}
    \label{fig:intelsat-colorpeak}
\end{figure*}

The phase light curves of IS10-02 (Figure \ref{fig:phasecurves-intelsat}) show a marked change in character before and after the docking with MEV-2. Figure \ref{fig:intelsat-colorpeak} (bottom panel) presents these light curves in greater detail, and includes additional observations obtained in 2018 and 2024 that confirm that these do indeed correspond to a long-term change in optical signature.

The 2018 observation was assembled from GEO survey-mode observations from the original SuperWASP North telescope \citep{Pollacco2006, Chote2019}. Sequences of $3 \times 10$\,s exposures were acquired every 20 minutes as part of a survey spanning the visible GEO belt. A small offset in the sampling phase each night allows a near-complete light curve to be assembled by combining three successive nights of observation, minimizing the impact of the solar declination change. The 2024 observation was obtained with the STING telescope \citep{Airey2025} as part of a follow-up survey of the \emph{PHANTOM ECHOES 2} field. Both instruments' data were reduced following the same procedures as for WTT.

Before docking, the light curve was dominated by a double-peaked glint at a phase angle of $\approx30^\circ$. A description of the Eurostar 3000's Attitude and Orbital Control Subsystem (AOCS) \citep{Martin2000} explains that both features (the double peak and the offset) are consequences of the AOCS, which adjusts the positioning of the solar arrays to manage torques imparted by solar radiation pressure. The mean solar array offset ($15^\circ$ is implied by the $30^\circ$ phase offset) controls the long term movements, and in particular allows for the offloading of torque from the momentum wheels. The $\pm2^\circ$ tilt between the two solar arrays (measured from the peak separation) has the effect of damping short term nutation effects.

After docking, MEV-2 takes over all attitude control for the combined system, so it should not be surprising that the glint shifts to a phase angle $\sim5^\circ$, and changes shape.

One particularly interesting operational detail is that the solar arrays on MEV-2 are oriented in the east-west direction to avoid being shadowed by IS10-02, which means that they cannot track the sun. They should therefore behave similarly to other `body' features in phase angle light curves. The post-docking glint shape shows a central peak with broad asymmetric shoulders: if this corresponds to the MEV-2 arrays (central peak) superimposed with IS10-02's arrays (shoulders) then we should expect to see the shape of this feature change with solar declination as the illumination angles will be different for the two sets of arrays. The 2024 observation appears to support this hypothesis, which will be explored further in a future study.

GOTO color observations were obtained targeting these glints before and after docking (Figure \ref{fig:intelsat-colorpeak}; top panels). The relative color shows no significant change ($\lesssim 0.05\,$mag) pre-docking, but the post-docking light curve suggests a small ($\approx 0.2\,$mag) redward shift during the central peak of the glint. This is consistent with Figure \ref{fig:goto-colors} showing that MEV-2 appears redder than IS10-02.

\begin{figure*}
    \centering
    \includegraphics{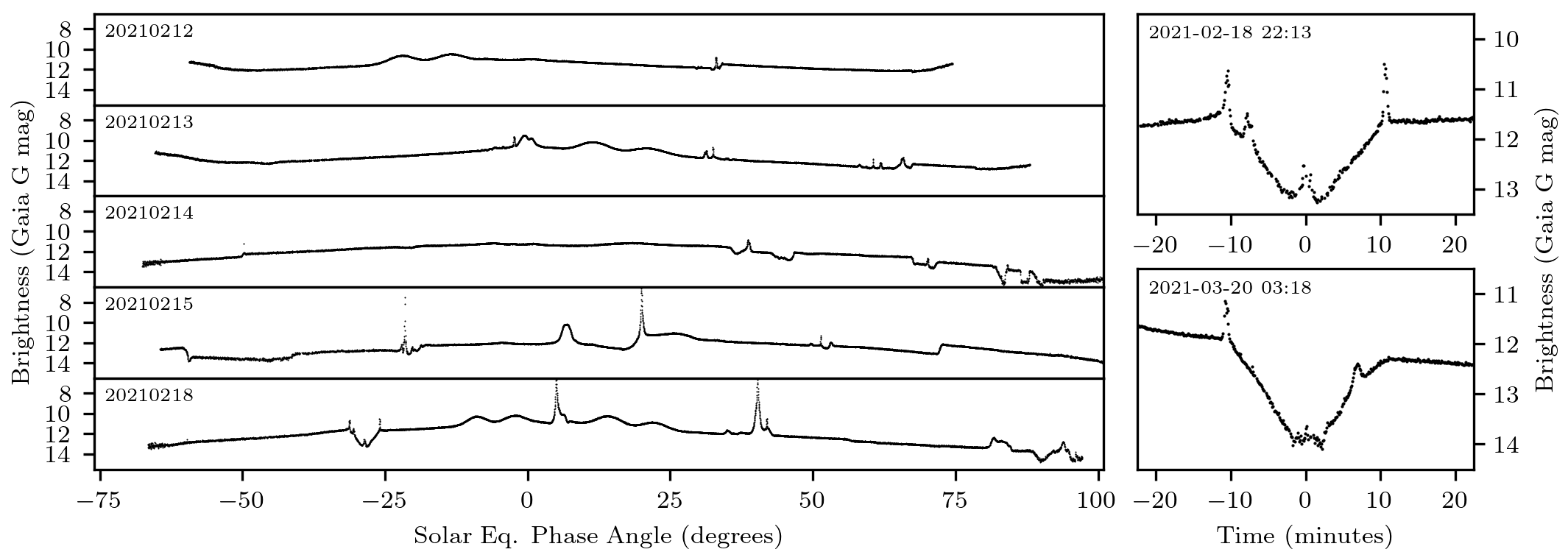}
    \caption{\textbf{Left:~}Observations of MEV-2 over five nights display very different behaviors, due presumably to the active maneuvering of MEV-2 around IS10-02. \textbf{Right:~}Close up views of what appears to be the same type of maneuver executed on two different nights.}
    \label{fig:mev2-maneuvers}
\end{figure*}

\subsection{MEV-2 Maneuvers}\label{sec:mev2-maneuvers}

MEV-2 performed extensive maneuvers and calibration activities in the weeks after it reached IS10-02 \citep{Pyrak2021}. These maneuvers produced clear signatures in the observed light curves. A subset of these are shown in Figure~\ref{fig:mev2-maneuvers}. Glints are seen to occur at arbitrary times/phase angles, as well as step changes in brightness that are indicative of attitude adjustments changing the relative angles between the Sun, MEV-2, and observer.

Of particular note were two features, shown inset in Figure~\ref{fig:mev2-maneuvers}, that appear to be mirror-images of each other. In both cases, the feature starts with a short ($\Delta t \lesssim1\,$min) glint, followed by a decrease in brightness of $\sim1.5\,$mag. Another short glint occurs in the middle of this feature, and the brightness then returns to a level consistent with the behavior before the glint. Both cases show a third glint on the larger-phase-angle side of the dip, and one of the nights shows a fourth glint immediately at the end of the feature.
The decrease in overall brightness suggest a temporary attitude change: either turning the solar panels away from their nominal pointing, a shadowing of the solar panels by the satellite body, or perhaps a combination of both. These features both occurred on nights immediately before MEV-2 moved into close proximity with IS10-02, so we infer that this signature is characteristic of the maneuvers performed while preparing for this close approach. 

Observations covering the first two weeks of March show no signs of maneuvers, so we speculate that MEV-2 was holding position while its operators analyzed data from its previous close approach.

\section{Future Work}

The work presented in this paper (in pre-published form) resulted in a successful funding bid for refurbishing and upgrading the SuperWASP North telescope into a dedicated wide-field multi-color RSO survey instrument. The upgraded instrument (now called STING) began routine operations in early 2023. A characterization of the instrument capabilities and initial survey results are presented in \cite{Airey2025}.
 
\section{Conclusions}
\label{sec:conclusions}

This study has demonstrated a variety of analyses that are enabled by high-quality photometric observations and, in particular, by supplementing a baseline photometric monitoring campaign with complementary snapshot observations from instrumentation that have different characteristics.

\begin{itemize}
\item Comparing optical positions versus passive RF highlights how critical it is for optical sensors to use high-quality absolute time-stamping and to account for any rolling shutter effects to avoid systematic errors. It also acts as a caution to distinguish between the precision of raw measurements (which is superb for passive RF) compared to the accuracy of any derived orbits (which may suffer from artifacts during fitting).

\item The phase-angle light curves of stabilized RSOs evolve over a timescale of several days due to the changing declination (latitude) angle of the sun. We presented a means to efficiently visualize these changes using a two-dimensional color map. These maps are sensitive to short timescale features that appear to be unique for each RSO, making them an effective optical signature for RSO characterization.

\item Precise measurements of relative color during a restricted range of phase angles are sufficient to uniquely distinguish between the six RSOs. There was no measurable change in the early-night color of IS10-02 before and after MEV-2 docked, but there are signs that the color becomes redder during the main solar panel glint; the time where MEV-2 is expected to provide its largest flux contribution to the combined light curve.

\item Simultaneous multi-color measurements of Thor 6 provide insight into its material composition, showing that one light curve feature is associated with a blue (solar-cell like) spectral profile, and another with a flat (silver MLI-like) profile.
 
\item Meteosat 11 was seen to dramatically increase in brightness and become bluer in the lead up to the March equinox, as expected for specular reflection from its cylindrical body. Its 0.6\,s spin period was detected during this bright state as a $\sim11\,$s alias signal.

\item Precise measurements of centroid positions during glints can be used to constrain the relative position of RSOs that are operating in close proximity.

\item Changes in the phase light curve glint signatures before and after docking show clear evidence for the change in operational state of IS10-02. Before docking, light curves show features associated with active attitude control. These features disappear after docking, when MEV-2 takes over attitude control for the combined vehicle stack.

\item Observations of MEV-2 during its initial RPO showed clear signatures indicating attitude changes. Periods of activity and inactivity provided insight into the operational behavior of the RSO.

\end{itemize}

The small-amplitude, short timescale brightness features seen in the phase-angle light curve maps are likely due to the specular reflection from flat surfaces that are inclined relative to the Sun-satellite-observer plane. If this assumption is correct, it is likely that these patterns would repeat on a yearly basis; this would be a powerful signature for identifying individual objects, and possibly satellite bus types. Future observations will be made to test this prediction.

Many of the results presented above were enabled by the careful application of techniques to correct for systematic errors and improve the photometric precision of the extracted photometry. We have described these in sufficient detail that others should be able to replicate similar corrections in their own pipelines.

\section*{Acknowledgements}
\label{sec:acknowledgements}

This work has made use of data obtained
using the Warwick CMOS test telescope operated on the island of La Palma by the University of Warwick in the Spanish Observatory del Roque de los Muchachos of the Instituto de Astrofísica de Canarias.

The Gravitational-wave Optical Transient Observer (GOTO) project acknowledges the support of the Monash-Warwick Alliance; University of Warwick; Monash University; University of Sheffield; University of Leicester; Armagh Observatory and Planetarium; the National Astronomical Research Institute of Thailand (NARIT); Instituto de Astrofísica de Canarias (IAC); University of Portsmouth; University of Turku; University of Manchester and the UK Science and Technology Facilities Council (STFC, grant numbers ST/T007184/1, ST/T003103/1 and ST/T000406/1).

This research was made possible through the use of the AAVSO Photometric All-Sky Survey (APASS), funded by the Robert Martin Ayers Sciences Fund and NSF AST-1412587.

The authors would like to thank SAFRAN Data Systems for providing passive RF data from their WeTrack\texttrademark~service, and in particular to Baptiste Guillot for providing technical support to our data analysis.

For the purpose of open access, the author has applied a Creative Commons Attribution (CC-BY) license to any Author Accepted Manuscript version arising from this submission.

\section*{Data availability}
The photometric data underlying this article will be shared on reasonable request to the corresponding author. Passive RF data were provided by SAFRAN Data Systems under license and cannot be shared without permission.

\appendix
\section{Observation Logs}
Details on the WTT and GOTO telescope data sets are provided in Tables \ref{tab:cmos-observations} and \ref{tab:goto-observations}.

\begin{table}[htbp]
    \centering
    \begin{tabular}{ccccc}
    Night    & Start Time & End Time & Exposure & Usable\\
    Starting & (UTC)      & (UTC)    & (s)      & Images\\
    \hline
2021-01-28 & 22:09 & 07:10 & 5    & 6121  \\
2021-01-29 & 21:15 & 07:15 & 5    & 6165  \\
2021-01-30 & 19:25 & 07:15 & 5    & 7758  \\
2021-01-31 & 19:25 & 07:15 & 5    & 8070  \\
2021-02-01 & 19:35 & 07:16 & 5    & 7676  \\
2021-02-02 & 19:30 & 07:17 & 5    & 7523  \\
2021-02-03 & 19:38 & 21:02 & 5    & 904   \\
2021-02-12 & 19:42 & 07:09 & 5    & 7792  \\
2021-02-13 & 19:36 & 07:09 & 5    & 7866  \\
2021-02-14 & 19:35 & 07:14 & 5    & 7944  \\
2021-02-15 & 19:49 & 07:14 & 5    & 7768  \\
2021-02-16 & 19:52 & 07:14 & 5    & 6066  \\
2021-02-18 & 19:41 & 07:09 & 5    & 7820  \\
2021-02-19 & 19:40 & 07:09 & 5    & 7822  \\
2021-02-22 & 19:44 & 07:04 & 5    & 7720  \\
2021-02-23 & 19:44 & 07:05 & 5    & 7730  \\
2021-03-01 & 19:48 & 06:58 & 5    & 7151  \\
2021-03-03 & 19:49 & 06:58 & 2.5  & 12186 \\
2021-03-09 & 19:53 & 23:07 & 5    & 1194  \\
2021-03-10 & 19:54 & 06:49 & 1, 5 & 10561 \\
2021-03-15 & 19:57 & 06:42 & 1, 5 & 11214 \\
2021-03-16 & 19:57 & 06:41 & 1, 5 & 11219 \\
2021-03-17 & 19:58 & 06:40 & 1, 5 & 11196 \\
2021-03-18 & 19:59 & 06:39 & 1, 5 & 11177 \\
2021-03-19 & 19:59 & 23:01 & 1, 5 & 9471  \\
2021-03-20 & 20:00 & 06:36 & 1, 5 & 11017 \\
2021-03-21 & 20:00 & 06:35 & 1, 5 & 11075 \\
2021-03-22 & 20:01 & 06:34 & 1, 5 & 11009 \\
2021-03-23 & 20:02 & 06:34 & 1, 5 & 10922 \\
2021-03-24 & 20:02 & 06:32 & 1, 5 & 10633 \\
2021-04-01 & 20:07 & 01:45 & 5    & 3849  \\
2021-04-02 & 20:08 & 06:21 & 5    & 6964  \\
2021-04-04 & 20:15 & 06:07 & 5    & 6906  \\
2021-04-14 & 20:15 & 06:07 & 5    & 2901  \\
2021-04-16 & 20:16 & 06:05 & 1, 5 & 7672  \\
2021-04-17 & 20:17 & 06:04 & 1, 5 & 8057  \\
2021-04-20 & 20:20 & 05:59 & 1, 5 & 8020  \\
2021-04-21 & 20:20 & 05:59 & 1, 5 & 7979  \\
2021-05-03 & 20:28 & 05:47 & 1, 5 & 6803  \\
2021-05-04 & 20:29 & 05:47 & 1, 5 & 7747  \\
2021-05-05 & 20:29 & 05:46 & 1, 5 & 7726  \\
2021-05-06 & 20:30 & 05:45 & 1, 5 & 7690
    \end{tabular}
    \caption{Observation log for the Warwick Test Telescope.}
    \label{tab:cmos-observations}
\end{table}

\begin{table}[htbp]
    \centering
    \begin{tabular}{ccccc}
    Night    & Start Time & End Time & Exposure & Usable\\
    Starting & (UTC)      & (UTC)    & (s)      & Images\\
    \hline
    2021-01-31 & 19:50 & 20:25 & 5 & 107 \\
    2021-02-01 & 19:55 & 20:28 & 5 & 112 \\
    2021-02-02 & 19:51 & 20:27 & 5 & 107 \\
    2021-02-12 & 19:58 & 20:34 & 5 & 107 \\
    2021-02-13 & 19:58 & 20:33 & 5 & 107 \\
    2021-02-14 & 19:47 & 20:23 & 5 & 107 \\
    2021-02-15 & 19:59 & 20:34 & 5 & 107 \\
    2021-02-16 & 20:02 & 20:38 & 5 & 107 \\
    2021-02-18 & 20:01 & 20:33 & 5 & 107 \\
    2021-02-19 & 20:03 & 20:35 & 5 & 107 \\
    2021-02-23 & 20:05 & 20:37 & 5 & 107 \\
    2021-03-03 & 20:16 & 20:51 & 5 & 107 \\
    2021-03-15 & 20:20 & 20:55 & 5 & 107 \\
    2021-03-21 & 20:24 & 20:58 & 5 & 107 \\
    2021-03-22 & 20:24 & 20:59 & 5 & 107 \\
    2021-03-22 & 01:52 & 02:23 & 1 & 119 \\
    2021-04-17 & 20:39 & 21:21 & 5 & 125 \\
    2021-04-17 & 23:58 & 00:36 & 1 & 149 \\
    2021-04-20 & 20:46 & 21:27 & 5 & 125 \\
    2021-04-21 & 20:36 & 21:17 & 5 & 125 \\
    2021-05-03 & 20:51 & 21:26 & 5 & 107 \\
    2021-05-04 & 20:53 & 21:28 & 5 & 107 \\
    \end{tabular}
    \caption{Observation log for the GOTO-North prototype.}
    \label{tab:goto-observations}
\end{table}

\bibliographystyle{jasr-model5-names}
\biboptions{authoryear}

\newcommand{\aap}[1]{\rm{A\&A}}
\newcommand{\aaps}[1]{\rm{A\&AS}}
\newcommand{\aj}[1]{\rm{AJ}}
\newcommand{\apj}[1]{\rm{ApJ}}
\newcommand{\mnras}[1]{\rm{MNRAS}}
\newcommand{\pasp}[1]{\rm{PASP}}

\bibliography{references}
\end{document}